\documentclass[]{aa}
\usepackage{natbib}         
\usepackage{graphicx}
\usepackage{longtable}
\usepackage[switch]{lineno} 
\linenumbers\modulolinenumbers[5]
\bibpunct{(}{)}{;}{a}{}{,} 

\def\coefRG{0.038} 
\def\coefMS{0.57}
\def\ndonnees{94} \def\nmodeles{43}

\newcommand{\ind}[1]{_{\mathrm{#1}}}
\newcommand{\diff}{\mathrm{d}}

\def\Kepler{\emph{Kepler}}

\newcommand{\refeq}[1]{(\ref{#1})}

\def\numax{\nu\ind{max}}
\def\nmax{n\ind{max}}
\newcommand{\np}{n}
\def\Dnu{\Delta\nu}
\def\Dnuobs{\Delta\nu\ind{obs}}
\def\epsobs{\varepsilon\ind{obs}}
\def\coefa{a}
\def\coefaMS{\coefa{}\ind{MS}}
\def\coefaRG{\coefa{}\ind{RG}}
\def\alfa{\alpha\ind{obs}}
\def\alfaRG{{\alfa}{}\ind{,RG}}
\def\alfaMS{{\alfa}{}\ind{,MS}}
\def\Dnuas{\Delta\nu\ind{as}}
\def\epsas{\varepsilon\ind{as}}
\def\Aas{A\ind{as}}

\def\ng{n\ind{g}}
\def\Ng{N\ind{g}}

\def\Teff{T\ind{eff}}

\def\Robs{R\ind{obs}}
\def\Mobs{M\ind{obs}}
\def\Ras{R\ind{as}}
\def\Mas{M\ind{as}}

\def\ng{n\ind{g}}

\newcommand\iref{_\odot}
\newcommand\dnus{{\Dnu}\iref}
\newcommand\numaxs{\numax{}_{,}{}\iref}
\newcommand\dnusas{\Dnu\ind{ref}}
\newcommand\numaxas{\nu\ind{ref}}
\newcommand\Ts{T\iref}
\newcommand\Rs{R\iref}
\newcommand\Ms{M\iref}
\newcommand\Rsis{R} 
\newcommand\Msis{M} 
\newcommand\Rmod{R\ind{mod}}
\newcommand\Mmod{M\ind{mod}}

\def\aCenA{$\alpha$\,Cen\,A} \def\aCenB{$\alpha$\,Cen\,B}
\def\muArae{$\mu$Arae}

\begin{document}
\title{Asymptotic and measured large frequency separations}
\titlerunning{Asymptotic and measured large frequency separations}
\author{B. Mosser\inst{1}\and
E. Michel\inst{1}\and
K. Belkacem\inst{1}\and
M.J. Goupil\inst{1}\and
A. Baglin\inst{1}\and
C. Barban\inst{1}\and
J. Provost\inst{2}\and
R. Samadi\inst{1}\and
M. Auvergne\inst{1}\and
C. Catala\inst{1}
} \offprints{B. Mosser}

\institute{LESIA, CNRS, Universit\'e Pierre et Marie Curie, Universit\'e Denis Diderot,
Observatoire de Paris, 92195 Meudon cedex, France; \email{benoit.mosser@obspm.fr}
\and Universit\'e de Nice-Sophia Antipolis, CNRS UMR 7293, Observatoire de la C\^ote d'Azur, Laboratoire J.L. Lagrange, BP 4229,
06304 Nice Cedex 04, France
}

\abstract{With the space-borne missions CoRoT and \emph{Kepler}, a
large amount of asteroseismic data is now available and has led to
a variety of work. So-called global oscillation parameters are
inferred to characterize the large sets of stars, perform ensemble
asteroseismology, and derive scaling relations. The mean large
separation is such a key parameter, easily deduced from the
radial-frequency differences in the observed oscillation spectrum
and closely related to the mean stellar density. It is therefore
crucial to measure it with the highest accuracy in order to obtain
the most precise asteroseismic indices.}
{As the conditions of measurement of the large separation do not
coincide with its theoretical definition, we revisit the
asymptotic expressions used for analyzing the observed oscillation
spectra. Then, we examine the consequence of the difference
between the observed and asymptotic values of the mean large
separation.}
{The analysis is focused on radial modes. We use series of
radial-mode frequencies in published analyses of stars with
solar-like oscillations to compare the asymptotic and
observational values of the large separation. This comparison
relies on the proper use of the second-order asymptotic
expansion.}
{We propose a simple formulation to correct the observed value of
the large separation and then derive its asymptotic counterpart.
The measurement of the curvature of the radial ridges in the
\'echelle diagram provides the correcting factor. We prove that,
apart from glitches due to stellar structure discontinuities, the
asymptotic expansion is valid from main-sequence stars to red
giants. Our model shows that the asymptotic offset is close to
1/4, as in the theoretical development, for low-mass,
main-sequence stars, subgiants and red giants.}
{High-quality solar-like oscillation spectra derived from precise
photometric measurements are definitely better described with the
second-order asymptotic expansion. The second-order term is
responsible for the curvature observed in the \'echelle diagrams
used for analyzing the oscillation spectra, and this curvature is
responsible for the difference between the observed and asymptotic
values of the large separation. Taking it into account yields a
revision of the scaling relations, which provides more accurate
asteroseismic estimates of the stellar mass and radius. After
correction of the bias (6\,\% for the stellar radius and 3\,\% for
the mass), the performance of the calibrated relation is about
4\,\% and 8\,\% for estimating, respectively, the stellar radius
and the stellar mass for masses less than $1.3\,M_\odot$; the
accuracy is twice as bad for higher mass stars and red giants.}


\maketitle

\voffset = 1.2cm

\section{Introduction\label{introduction}}

The amount of asteroseismic data provided by the space-borne
missions CoRoT and \Kepler\ has given rise to ensemble
asteroseismology. With hundreds of stars observed from the main
sequence to the red giant branch, it is possible to study
evolutionary sequences and to derive seismic indices from global
seismic observational parameters. These global parameters can be
measured prior to any complete determination of the individual
mode frequencies and are able to provide global information on the
oscillation spectra \citep[e.g.,][]{2008Sci...322..558M}. For
instance, the mean large separation $\Dnu$ between consecutive
radial-mode frequencies and the frequency $\numax$ of maximum
oscillating signal are widely used to provide estimates of the
stellar mass and radius
\citep[e.g.,][]{2010A&A...522A...1K,2010A&A...517A..22M}. Almost
all other scaling relations make use of $\Dnu$, as, for instance,
the scaling reations governing the amplitude of the oscillation
signal \citep[e.g.,][]{2011ApJ...737L..10S,2012A&A...537A..30M}.
In case an oscillation spectrum is determined with a low
signal-to-noise ratio, the mean large separation is the single
parameter that can be precisely measured
\citep[e.g.,][]{2001ApJ...549L.105B,2009A&A...506...33M,2010A&A...524A..47G}.

The definition of the large separation relies on the asymptotic
theory, valid for large values of the eigenfrequencies,
corresponding to large values of the radial order. However, its
measurement is derived from the largest peaks seen in the
oscillation spectrum in the frequency range surrounding $\numax$,
thus in conditions that do not directly correspond to the
asymptotic relation. This difference can be taken into account in
comparisons with models for a specific star, but not in the
consideration of large sets of stars for statistical studies.

In asteroseismology, ground-based observations with a limited
frequency resolution have led to the use of simplified and
incomplete forms of the asymptotic expansion. With CoRoT and
\Kepler\ data, these simplified forms are still in use, but
observed uncertainties on frequencies are much reduced. As for the
Sun, one may question using the asymptotic expansion, since the
quality of the data usually allows one to go beyond this
approximate relation. Moreover, modeling has shown that acoustic
glitches due to discontinuities or important gradients in the
stellar structure may hamper the use of the asymptotic expansion
\citep[e.g.,][]{1994A&A...282...73A}.

In this work, we aim to use the asymptotic relation in a proper
way to derive the generic properties of a solar-like oscillation
spectrum. It seems therefore necessary to revisit the different
forms of asymptotic expansion since $\Dnu$ is introduced by the
asymptotic relation. We  first show that it is necessary to use
the asymptotic expansion including its second-order term, without
simplification compared to the theoretical expansion. Then, we
investigate the consequence of measuring the large separation at
$\numax$ and not in asymptotic conditions. The relation between
the two values, $\Dnuas$ for the asymptotic value and $\Dnuobs$
for the observed one, is developed in Section \ref{lien}. The
analysis based on radial-mode frequencies found in the literature
is carried out in Section \ref{analysis} to quantify the relation
between the observed and asymptotic parameters, under the
assumption that the second-order asymptotic expansion is valid for
describing the radial oscillation spectra. We verify that this
hypothesis is valid when we consider two regimes, depending on the
stellar evolutionary status. We discuss in Section
\ref{discussion} the consequences of the relation between the
observed and asymptotic parameters. We also use the comparison of
the seismic and modeled values of the stellar mass and radius to
revise the scaling relations providing $\Msis$ and $\Rsis$
estimates.

\begin{figure}
\includegraphics[width=8.8cm]{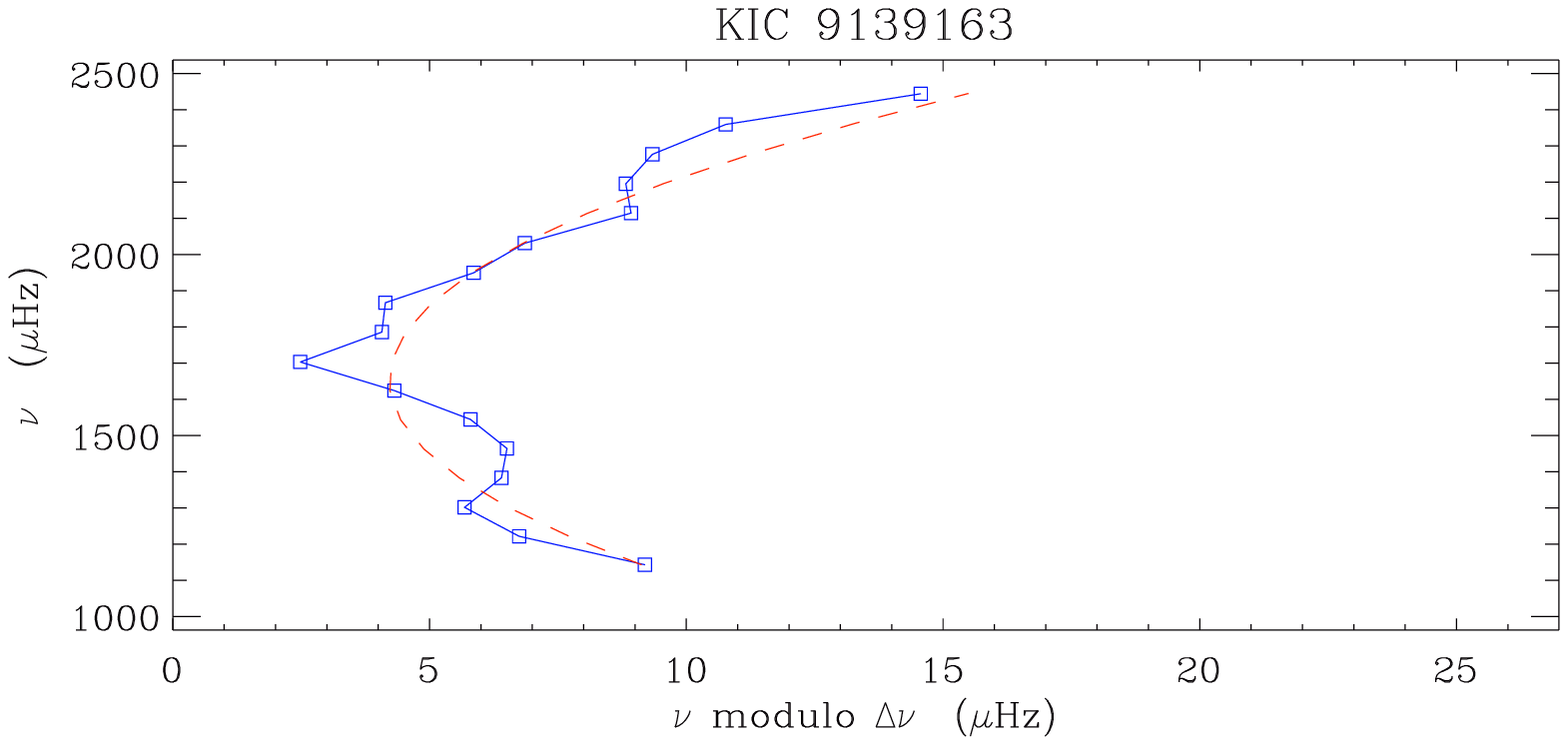}
\caption{\'Echelle diagram of the radial modes of the star KIC
9139163 \citep[from][]{2012A&A...543A..54A}. The red dashed line
indicates the quadratic fit that mimics the curvature.
\label{fig0}}
\end{figure}
\begin{figure}
\includegraphics[width=8.8cm]{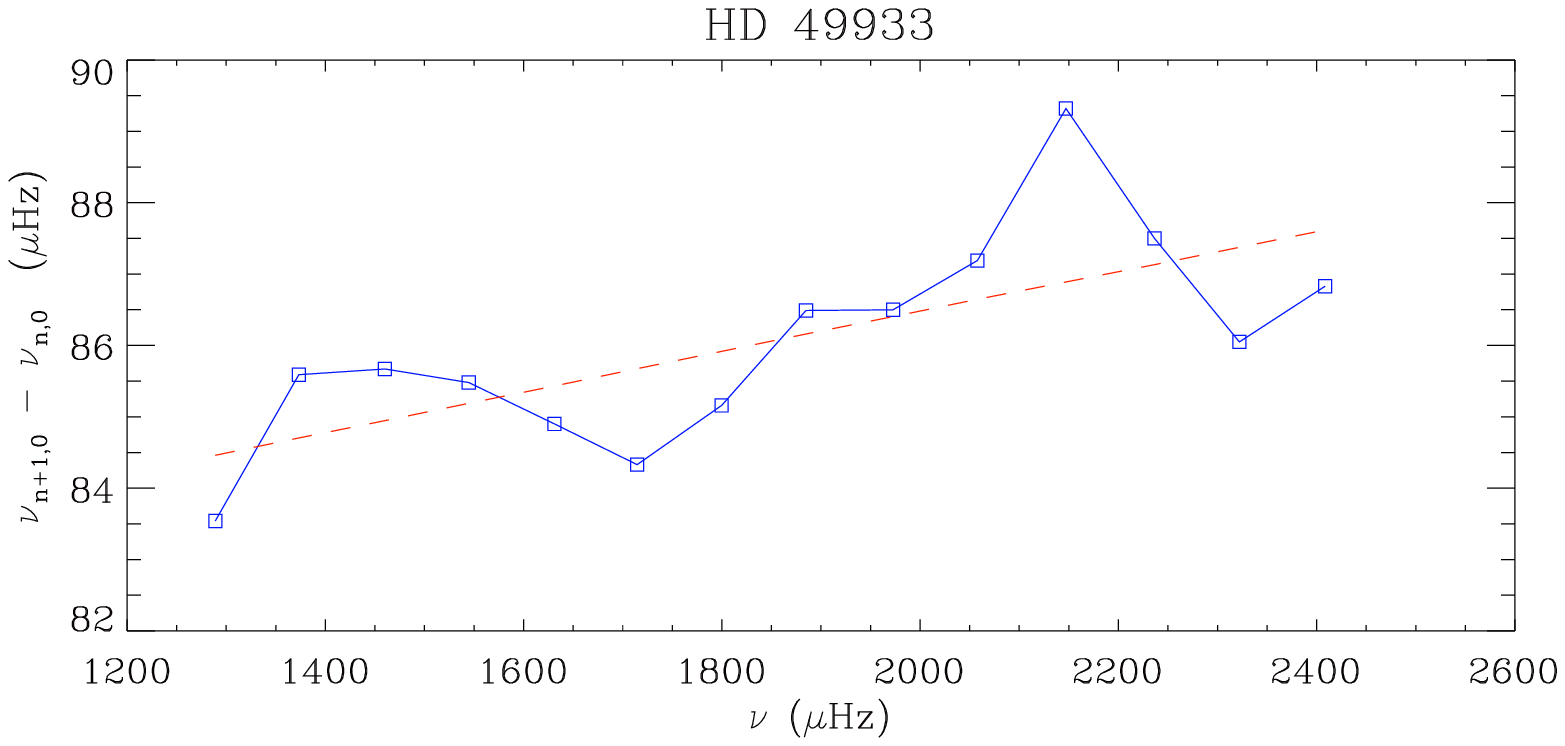}
\caption{Variation of the large separations $\nu_{n+1,0} -
\nu_{n,0}$ as a function of $\nu_{n,0}$ for the star HD\,49933
\citep[from][]{2009A&A...507L..13B}. The red dashed line indicates
a linear fit. \label{fig1}}
\end{figure}

\section{Asymptotic relation versus observed parameters\label{lien}}

\subsection{The original asymptotic expression}

The oscillation pattern of low-degree oscillation pressure modes
can be described by a second-order relation \citep[Eqs. 65-74
of][]{1980ApJS...43..469T}. This approximate relation is called
asymptotic, since its derivation is strictly valid only for large
radial orders. The development of the eigenfrequency
$\nu_{\np,\ell}$ proposed by Tassoul includes a second-order term,
namely, a contribution in $1/\nu_{\np,\ell}$:
\begin{equation}\label{asymp_ori}
\nu_{\np,\ell} = \left(\np  + {\ell\over 2} + \varepsilon \right)
\; \Dnu - \left[\ell (\ell+1)\; d_0 + d_1\right] {\Dnu^2 \over
\nu_{\np,\ell}}
\end{equation}
where $\np$ is the p-mode radial order, $\ell$ is the angular
degree, $\Dnu$ is the large separation, and $\varepsilon$ is a
constant term. The terms $\Dnu$ and $\varepsilon$ are discussed
later; the dimensionless term $d_0$ is related to the gradient of
sound speed integrated over the stellar interior; $d_1$ has a
complex form.

\subsection{The asymptotic expansion used in practice}

When reading the abundant literature on asteroseismic
observations, the most common forms of the asymptotic expansion
used for interpreting observed low-degree oscillation spectra
\citep[e.g.,][for ground-based
observations]{1991A&A...251..356M,2001ApJ...549L.105B,2005A&A...440..609B}
are similar to the approximate form
\begin{equation}\label{asymp2}
\nu_{\np,\ell} \simeq \left(\np  + {\ell\over 2} + \varepsilon_0
\right) \; \Dnu_0 - \ell (\ell+1)\; D_0 .
\end{equation}
Compared to Eq.~\refeq{asymp_ori}, the contribution of $d_1$ and
the variation of the denominator varying as $\nu_{n,\ell}$ are
both omitted. This omission derives from the fact that
ground-based observations have a too coarse frequency resolution.
Equation \refeq{asymp2} is still in use for space-borne
observations that provide a much better frequency resolution
\citep[e.g.,][]{2011A&A...534A...6C,2012ApJ...751L..36W,2012ApJ...757..190C}.
The large separation $\Dnu_0$ and the offset $\varepsilon_0$ are
supposed to play the roles of $\Dnu$ and $\varepsilon$ in
Eq.~\refeq{asymp_ori}: this is not strictly exact, as shown later.

Observationally, the terms $\Dnu_0$ and $D_0$ can be derived from
the mean frequency differences
\begin{eqnarray}
  \Dnu_0 & = & \langle \nu_{\np+1,\ell} - \nu_{\np,\ell} \rangle, \label{Dnu0}\\
  D_0    & = & \langle {1 \over 4\ell+6}\, (\nu_{\np,\ell} -
  \nu_{\np-1,\ell+2}) \rangle,
\end{eqnarray}
where the square brackets represent the mean values in the
observed frequency range. For radial modes, the asymptotic
expansion reduces to
\begin{equation}\label{tassoul_alt}
\nu_{\np,0} = \left(\np  + \varepsilon_0\right) \; \Dnu_0.
\end{equation}
Clearly, this form cannot account for an accurate description of
the radial-mode pattern, since the \'echelle diagram
representation shows an noticeable curvature for most stars with
solar-like oscillations at all evolutionary stages. This
curvature, always with the same concavity sign, is particularly
visible in red giant oscillations
\citep[e.g.,][]{2011A&A...525L...9M,2012A&A...541A..51K}, which
show solar-like oscillations at low radial order
\citep[e.g.,][]{2009Natur.459..398D,2010ApJ...713L.176B}.
Figure~\ref{fig0} shows a typical example of the non-negligible
curvature in the \'echelle diagram of a main-sequence star with
many radial orders. The concavity corresponds equivalently to a
positive gradient of the large separation (Fig.~\ref{fig1}). This
gradient is not reproduced by Eq.~\refeq{tassoul_alt}, stressing
that this form of the asymptotic expansion is not adequate for
reporting the global properties of the radial modes observed with
enough precision. Hence, it is necessary to revisit the use of the
asymptotic expression for interpreting observations and to better
account for the second-order term.

\subsection{Radial modes depicted by second-order asymptotic expansion}

We have chosen to restrict the analysis to radial modes.  We
express the second-order term varying in $\nu^{-1}$ of
Eq.~\refeq{asymp_ori} with a contribution in $n^{-1}$:
\begin{equation}\label{tassoul_as}
\nu_{\np,0} = \left(\np  + \epsas + {\Aas\over \np } \right) \;
\Dnuas .
\end{equation}
Compared to Eq.~\refeq{asymp_ori}, we added the subscript $as$ to
the different terms in order to make clear that, contrary to
Eq.~\refeq{asymp2}, we respect the asymptotic condition. We
replaced the second-order term in $1/\nu$ by a term in $1/n$
instead of $1/(n+\epsas)$, since the contribution of $\epsas$ in
the denominator can be considered as a third-order term in $n$.

The large separation $\Dnuas$ is related to the stellar acoustic
diameter by
\begin{equation}\label{acoustic}
\Dnuas = \left(2\int_0^R {\diff r\over c}\right)^{-1},
\end{equation}
where $c$ is the sound speed. We note that $\Dnu_0$, different
from the asymptotic value $\Dnuas$, cannot directly provide the
integral value of $1/c$. We also note that the offset $\epsas$ has
a fixed value in the original work of Tassoul:
\begin{equation}\label{surface}
\epsas{}\ind{,Tassoul} =  {1 \over 4} .
\end{equation}
In the literature, one also finds that $\epsas = 1/4 + a(\nu)$,
where $a(\nu)$ is determined by the properties of the near-surface
region \citep{1992MNRAS.257...62C}.

The second-order expansion is valid for large radial orders only,
when the second-order term $\Aas/n$ is small. This means that the
large separation $\Dnuas$ corresponds to the frequency difference
between radial modes at high frequency only, but not at $\numax$.
We note that the second-order term can account for the curvature
of the radial ridge in the \'echelle diagram, with a positive
value of $\Aas$ for reproducing the sign of the observed
concavity.

\begin{figure}
\includegraphics[width=8.8cm]{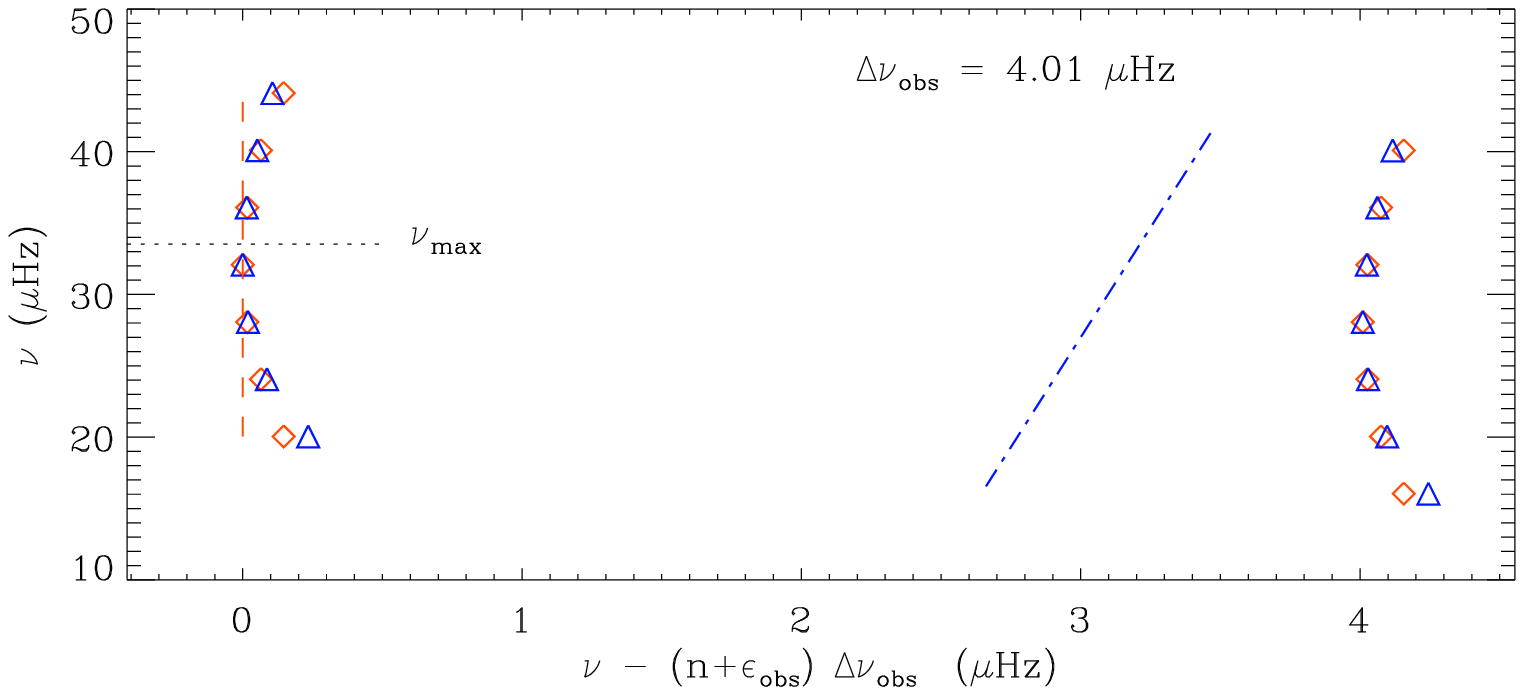}
\includegraphics[width=8.8cm]{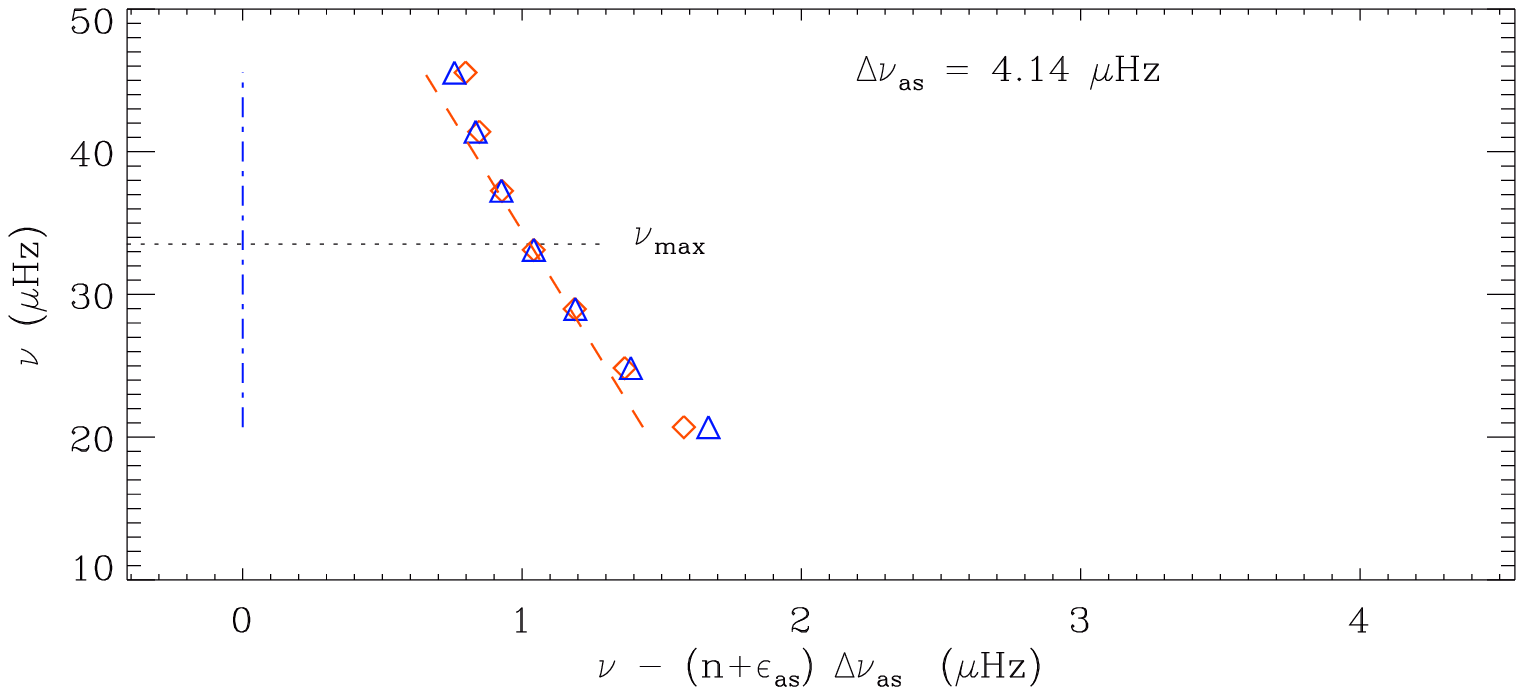}
\caption{\'Echelle diagrams of the radial modes of a typical
red-clump giant, comparing the asymptotic expansion
(Eq.~\refeq{tassoul_as}, blue triangles) and the development
describing the curvature (Eq.~\ref{tassoul_obs}, red diamonds).
{\sl Top:} diagram based on $\Dnuobs$ observed at $\numax$; the
dashed line indicates the vertical asymptotic line at $\numax$;
the dot-dashed line indicates the asymptotic line at high
frequency. For clarity, the ridge has been duplicated modulo
$\Dnuobs$. {\sl Bottom:} diagram based on $\Dnuas$; the dot-dashed
line indicates the vertical asymptotic line at high frequency.
\label{asymp_ech1}}
\end{figure}

\subsection{Taking into account the curvature}

In practice, the large separation is necessarily obtained from the
radial modes with the largest amplitudes observed in the
oscillation pattern around $\numax$
\citep[e.g.,][]{2009A&A...508..877M}. In order to reconcile
observations at $\numax$ and asymptotic expansion at large
frequency, we must first consider the curvature of the ridge. To
enhance the quality of the fit of radial modes in red giants,
\cite{2011A&A...525L...9M} have proposed including the curvature
of the radial ridge with the expression
\begin{equation}\label{tassoul_obs}
\nu_{\np,0} = \left(\np  + \epsobs
 + {\alfa \over 2}\; [ \np - \nmax ]^2 \right) \Dnuobs,
\end{equation}
where $\Dnuobs$ is the observed large separation, measured in a
wide frequency range around the frequency $\numax$ of maximum
oscillation amplitude, $\alfa$ is the curvature term, and
$\epsobs$ is the offset. We have also introduced the dimensionless
value of $\numax$, defined by $\nmax= \numax / \Dnuobs$. Similar
fits have already been proposed for the oscillation spectra of the
Sun
\citep[][]{1983SoPh...82..469C,1983SoPh...82...55G,1983SoPh...82...75S},
of \aCenA\ \citep{2004ApJ...614..380B}, \aCenB\
\citep{2005ApJ...635.1281K}, HD\,203608
\citep{2008A&A...488..635M}, Procyon \citep{2008A&A...478..197M},
and HD\,46375 \citep{2010A&A...524A..47G}. In fact, such a fit
mimics a second-order term and provides a linear gradient in large
separation:
\begin{equation}\label{gradient}
{\nu_{\np+1,0} - \nu_{\np-1,0}\over 2} = \left( 1 + \alfa \left[ \np - \nmax \right] \right) \Dnuobs.
\end{equation}
The introduction of the curvature may be considered as an
empirical form of the asymptotic relation. It reproduces the
second-order term of the asymptotic expansion, which has been
neglected in Eq.~\refeq{asymp2}, with an unequivocal
correspondence. It relies on a global description of the
oscillation spectrum, which considers that the mean values of the
seismic parameters are determined in a large frequency range
around $\numax$ \citep{2009A&A...508..877M}. Such a description
has shown interesting properties when compared to a local one that
provides the large separation from a limited frequency range only
around $\numax$ \citep{2011MNRAS.415.3539V,2012A&A...544A..90H}.

It is straightforward to make the link between both asymptotic and
observed descriptions of the radial oscillation pattern with a
second-order development in $(\np-\nmax)/\nmax$ of the asymptotic
expression. From the identification of the different orders in
Eqs.~\refeq{tassoul_as} and \refeq{tassoul_obs} (constant, varying
in $\np$ and in $\np^{2}$), we then get
\begin{eqnarray}
  \Dnuas &=& \Dnuobs\   \left({ 1 + {\nmax\alfa\over2}} \right),  \label{eqdnu}\\
  \Aas   &=& {\alfa \over 2} {\nmax^3 \over 1  + \nmax \displaystyle{\alfa\over 2}} , \label{eqsecond0}\\
  \epsas &=& {\epsobs - \nmax^2 \alfa \over 1  + \nmax \displaystyle{\alfa\over 2}} . \label{eqeps0}
\end{eqnarray}
When considering that the ridge curvature is small enough $(\nmax
\alfa/2 \ll 1)$, $\Aas$ and $\epsas$ become
\begin{eqnarray}
  \Aas   &\simeq& {\alfa \over 2}\ \nmax^3 , \label{eqsecond}\\
  \epsas &\simeq& \epsobs \left({1-{\nmax\alfa\over2}}\right) - \nmax^2 \alfa .\label{eqeps}
\end{eqnarray}
These developments provide a reasonable agreement with the
previous exact correspondence between the asymptotic and observed
forms.

The difference between the observed and asymptotic values of
$\varepsilon$ includes a systematic offset in addition to the
rescaling term $\left({1-{\nmax\alfa / 2}}\right)$. This comes
from the fact that the measurement of $\epsobs$ significantly
depends on the measurement of $\Dnuobs$: a relative change $\eta$
in the measurement of the large separation translates into an
absolute change of the order of $-\nmax \eta$. This indicates that
the measurement of $\epsobs$ is difficult since it includes large
uncertainties related to all effects that affect the measurement
of the large separation, such as structure discontinuities or
significant gradients of composition
\citep{2010A&A...520L...6M,2012A&A...540A..31M}.

\subsection{\'Echelle diagrams}

The \'echelle diagrams in Fig.~\ref{asymp_ech1} compare the radial
oscillation patterns folded with $\Dnuobs$ or $\Dnuas$. In
practice, the folding is naturally based on the observations of
quasi-vertical ridges and provides $\Dnuobs$ observed at $\numax$
(Fig.~\ref{asymp_ech1} top). A folding based on $\Dnuas$ would not
show any vertical ridge in the observed domain
(Fig.~\ref{asymp_ech1} bottom). In Fig.~\ref{asymp_ech1} we have
also compared the asymptotic spectrum based on the parameters
$\Dnuas$, $\Aas$, and $\epsas$ to the observed spectrum, obeying
Eq.~\refeq{tassoul_obs}. Perfect agreement is naturally met for
$\nu\simeq\numax$. Differences vary as $\alfa (n-\nmax)^3$. They
remain limited to a small fraction of the large separation, even
for the orders far from $\nmax$, so that the agreement of the
simplified expression is satisfactory in the frequency range where
modes have appreciable amplitudes. Comparison of the \'echelle
diagrams illustrates that the observable $\Dnuobs$  significantly
differs from the physically-grounded asymptotic value $\Dnuas$.

\addtocounter{table}{1}

\section{Data and analysis\label{analysis}}

\subsection{Observations: main-sequence stars and subgiants}

Published data allowed us to construct a table of observed values
of $\alfa$ and $\epsobs$ as a function of the observed large
separation $\Dnuobs$ (Table \ref{valeurs}, with \ndonnees\ stars).
We considered observations of subgiants and main-sequence stars
observed by CoRoT or by \Kepler. We also made use of ground-based
observations and solar data. All references are given in the
caption of Table \ref{valeurs}.

We have considered the radial eigenfrequencies, calculated local
large spacings $\nu_{n+1,0} -\nu_{n,0}$, and derived $\Dnuobs$ and
$\alfa$ from a linear fit corresponding to the linear gradient
given by Eq.~\refeq{gradient}. The term $\epsobs$ is then derived
from Eq.~\refeq{tassoul_obs}. With the help of Eqs.~\refeq{eqdnu},
\refeq{eqsecond0} and \refeq{eqeps0}, we analyzed the differences
between the observables $\Dnuobs$, $\epsobs$, $\alfa$ and their
asymptotic counterparts $\Dnuas$, $\epsas$, and $\Aas$. We have
chosen to express the variation with the parameter $\nmax$, rather
than $\Dnuobs$ or $\numax$. Subgiants have typically $\nmax \ge
15$, and main-sequence stars $\nmax \ge 18$.

\subsection{Observations: red giants}

We also considered observations of stars on the red giant branch
that have been modelled. They were analyzed exactly as the
less-evolved stars. However, this limited set of stars cannot
represent the diversity of the thousands of red giants already
analyzed in both the CoRoT and \Kepler\ fields
\citep[e.g.,][]{2009A&A...506..465H,2010ApJ...713L.176B,2010A&A...517A..22M,2010ApJ...713L.182S}.
We note, for instance, that their masses are higher than the mean
value of the largest sets. Furthermore, because the number of
excited radial modes is much more limited than for main-sequence
stars \citep{2010A&A...517A..22M}, the observed seismic parameters
suffer from a large spread. Therefore, we also made use of the
mean relation found for the curvature of the red giant radial
oscillation pattern \citep{2012A&A...548A..10M}
\begin{equation}
 \alfaRG = 0.015 \ \Dnuobs^{-0.32}. \label{alfaRG}
\end{equation}
This relation, when expressed as a function of $\nmax$ and taking
into account the scaling relation $\Dnu\propto \numax^{0.75}$
\citep[e.g.,][]{2011A&A...525A.131H}, gives $\alfaRG = 0.09 \
\nmax^{-0.96}$. The exponent of this scaling relation is close to
$-1$, so that for the following study we simply consider the fit
\begin{equation}\label{fit_alpha_geantes}
 \alfaRG =  2\,\coefaRG \, \nmax^{-1} ,
\end{equation}
with $\coefaRG = \coefRG \pm 0.002$. Red giants have typically
$\nmax \le 15$.

\begin{figure}
\includegraphics[width=8.8cm]{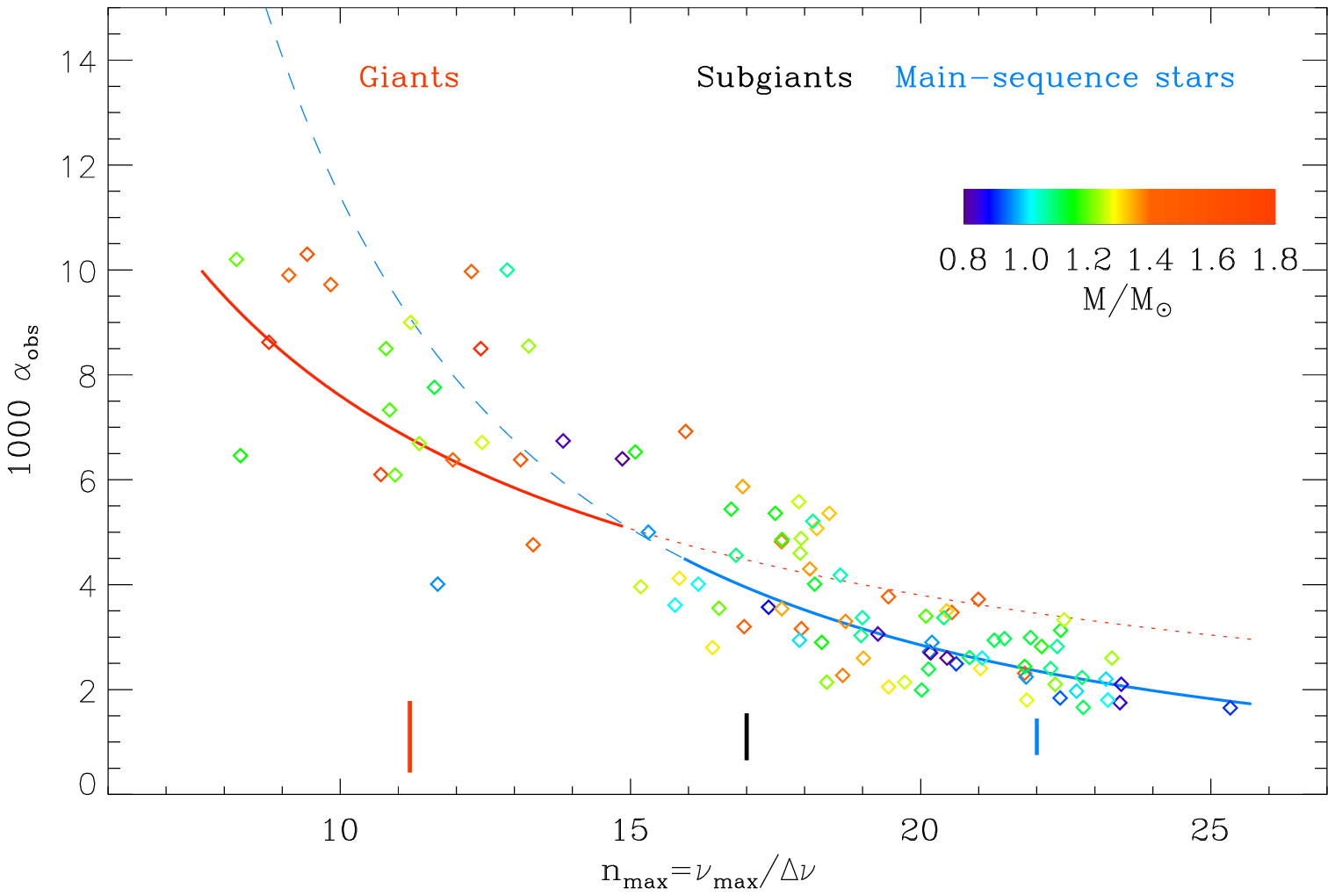}
\caption{Curvature $10^3\ \alfa$  as a function of $\nmax =
\numax/\Dnu$. The thick line corresponds to the fit in
$\nmax^{-1}$ established for red giants, and the dotted line to
its extrapolation towards larger $\nmax$. The dashed line provides
an acceptable fit in  $\nmax^{-2}$ valid for main-sequence stars.
Error bars indicate the typical 1-$\sigma$ uncertainties in three
different domains. The color code of the symbols provides an
estimate of the stellar mass; the colors of the lines correspond
to the different regimes.\label{asymp_second}}
\end{figure}

\begin{figure}
\includegraphics[width=8.8cm]{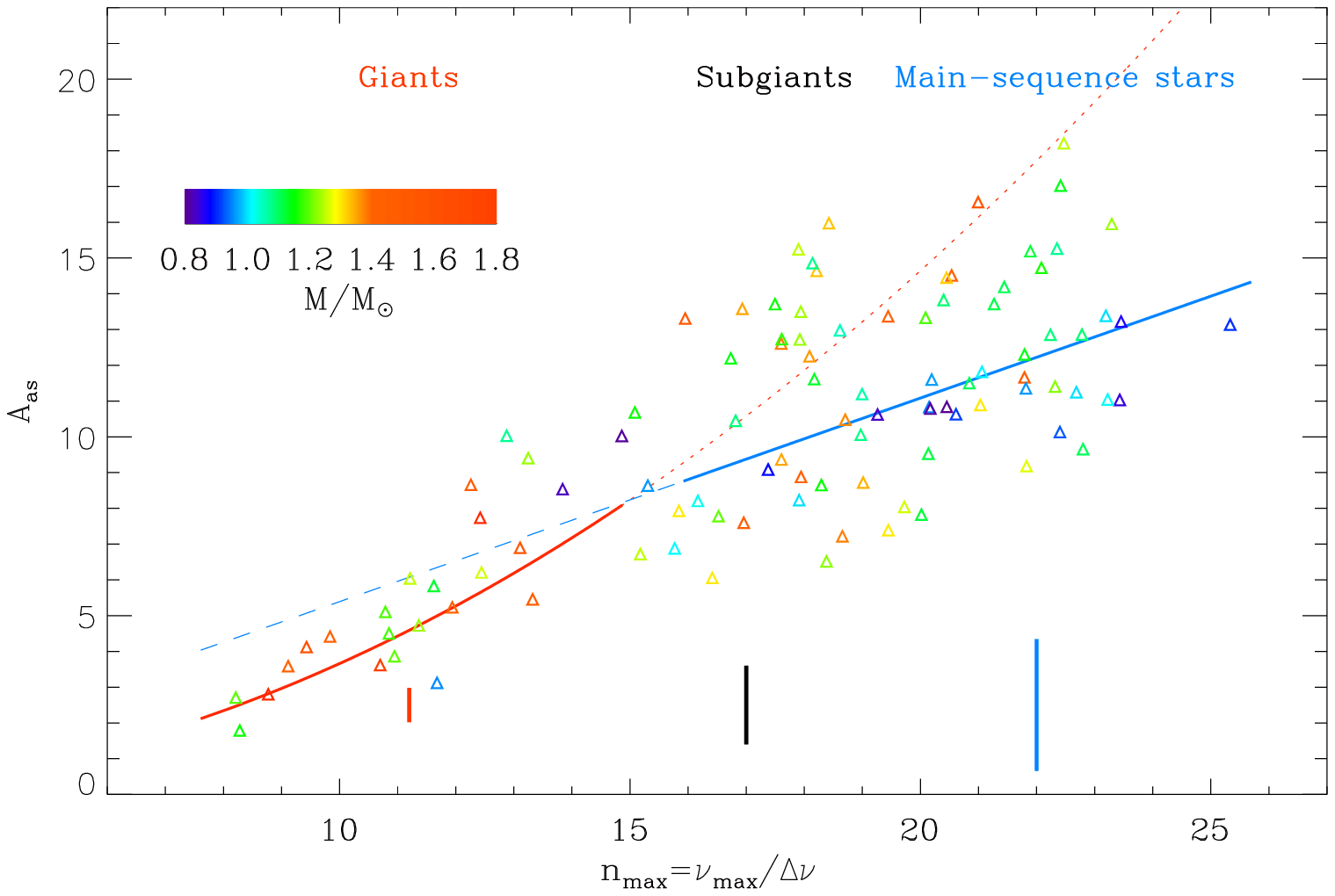}
\caption{Same as Fig.~\ref{asymp_second}, for the second-order
asymptotic term $\Aas$ as a function of $\nmax$. The fits of
$\Aas$ in the different regimes are derived from the fits of
$\alfa$ in Fig.~\ref{asymp_second} and the relation provided by
Eq.~\refeq{eqsecond0}. \label{asymp_A}}
\end{figure}

\begin{figure}
\includegraphics[width=8.8cm]{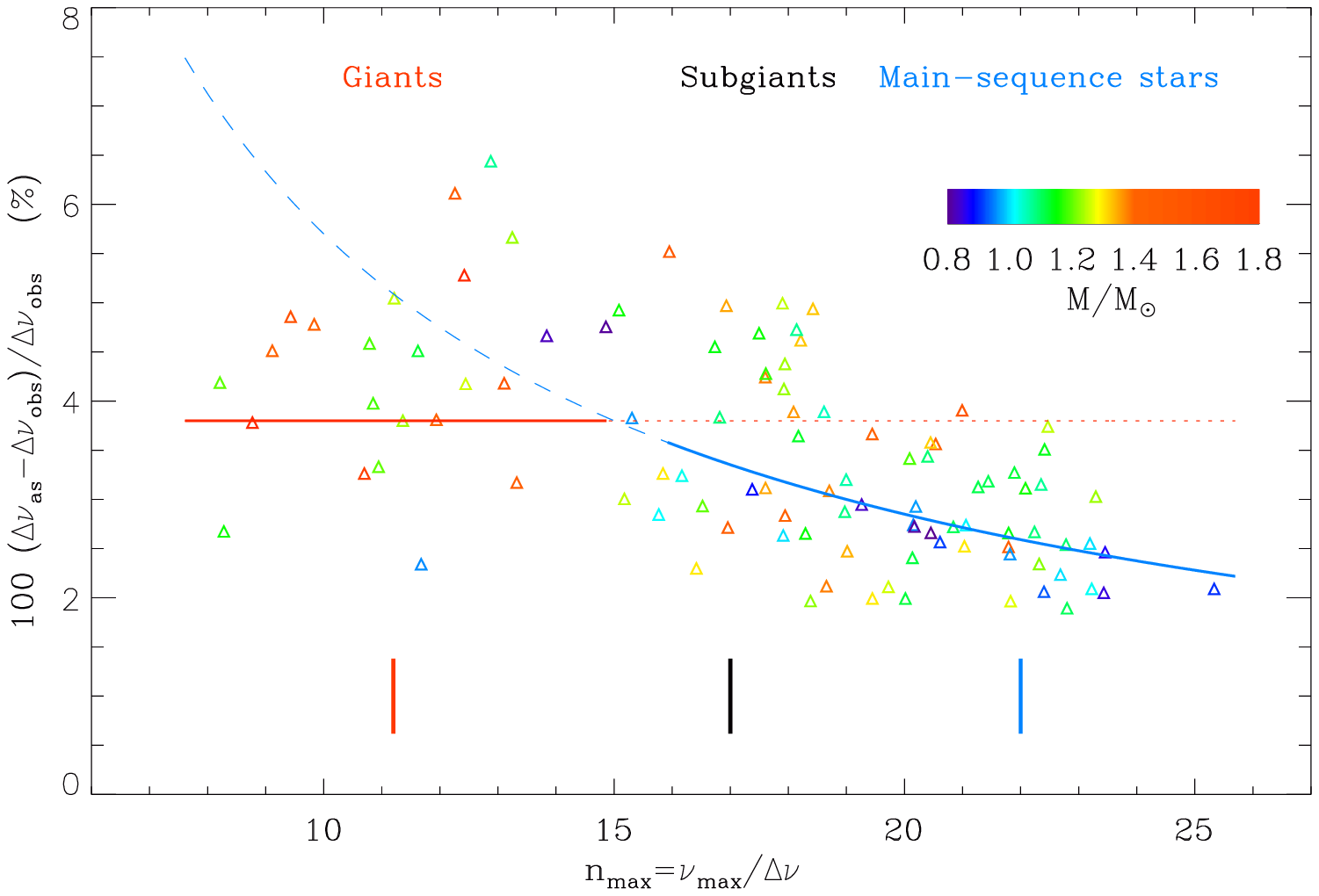}
\caption{Same as Fig.~\ref{asymp_second}, for the relative
difference of the large separations, equivalent to $\nmax\alfa/2$,
as a function of $\nmax$. \label{asymp_dnu}}
\end{figure}

\subsection{Second-order term $\Aas$ and curvature $\alfa$}

We first examined the curvature $\alfa$ as a function of $\nmax$
(Fig.~\ref{asymp_second}), since this term governs the relation
between the asymptotic and observed values of the large separation
(Eq.~\refeq{eqdnu}). A large spread is observed because of the
acoustic glitches caused by structure discontinuities
\citep[e.g.,][]{2010A&A...520L...6M}. Typical uncertainties of
$\alfa$ are about 20\,\%. We are interested in the mean variation
of the observed and asymptotic parameters, so that the glitches
are first neglected and later considered in Section
\ref{limitation}. The large spread of the data implies that, as is
well known, the asymptotic expansion cannot precisely relate all
of the features of a solar-like oscillation spectrum. However,
this spread does not invalidate the analysis of the mean evolution
of the observed and asymptotic parameters with frequency.

The fit of $\alfa$ derived from red giants, valid when $\nmax$ is
in the range [7, 15], does not hold for less-evolved stars with
larger $\nmax$  (Fig.~\ref{asymp_second}). It could reproduce part
of the observed curvature of subgiants but yields too large values
in the main-sequence domain. In order to fit main-sequence stars
and subgiants, it seems necessary to modify the exponent of the
relation $\alfa (\nmax)$. When restricted to main-sequence stars,
the fit of $\alfa (\nmax)$ provides an exponent of about
$-2\pm0.3$. We thus chose to fix the exponent to the integer value
$-2$:
\begin{equation}\label{fit_alpha_MS}
  \alfaMS = 2\,\coefaMS \, \nmax^{-2} ,
\end{equation}
with $\coefaMS = \coefMS \pm 0.02$. Having a different fit
compared to the red giant case (Eq.~\refeq{alfaRG}) is justified
in Section \ref{epsilon}: even if a global fit in $\nmax^{-1.5}$
should reconcile the two regimes, we keep the two regimes since we
also have to consider the fits of the other asymptotic parameters,
especially $\epsobs$. We also note a gradient in mass: low-mass
stars have systematically lower $\alfa$ than high-mass stars.
Masses were derived from the seismic estimates when modeled masses
are not available (Table~\ref{valeurs}). At this stage, it is
however impossible to take this mass dependence into
consideration.

As a consequence of Eq.~\refeq{eqsecond}, we find that the
second-order asymptotic term $\Aas$ scales as $\nmax^2$ for red
giants and as $\nmax$ for less-evolved stars (Fig.~\ref{asymp_A}).
We note, again, a large spread of the values, which is related to
the acoustic glitches.

\subsection{Large separations $\Dnuas$ and $\Dnuobs$}

The asymptotic and observed values of the large separations of the
set of stars are clearly distinct (Fig.~\ref{asymp_dnu}).
According to Eq.~\refeq{eqdnu}, the correction from $\Dnuobs$ to
$\Dnuas$ has the same relative uncertainty as $\alfa$. The
relative difference between $\Dnuas$ and $\Dnuobs$ increases when
$\nmax$ decreases and reaches a constant maximum value of about
4\,\% in the red giant regime, in agreement with
Eq.~\refeq{fit_alpha_geantes}. Their difference is reduced at high
frequency for subgiants and main-sequence stars, as in the solar
case, where high radial orders are observed, but still of the
order to 2\,\%. This relative difference, even if small, depends
on the frequency. This can be represented, for the different
regimes, by the fit
\begin{equation}\label{scadnu}
\Dnuas = \left( 1 + \zeta\right) \ \Dnuobs,
\end{equation}
with
\begin{eqnarray}
  \zeta &=& \displaystyle{\coefMS\over\nmax} \qquad\hbox{(main-sequence regime: }  \nmax \ge 15), \label{corM_MS}\\
  \zeta &=& \coefRG \qquad \ \ \hbox{(red giant regime: } \nmax \le 15). \label{corM_RGB}
\end{eqnarray}
The consequence of these relations is examined in
Section~\ref{scaling}. As for the curvature, the justification of
the two different regimes is based on the analysis of the offset
$\epsobs$.

\begin{figure*}
\includegraphics[width=16cm]{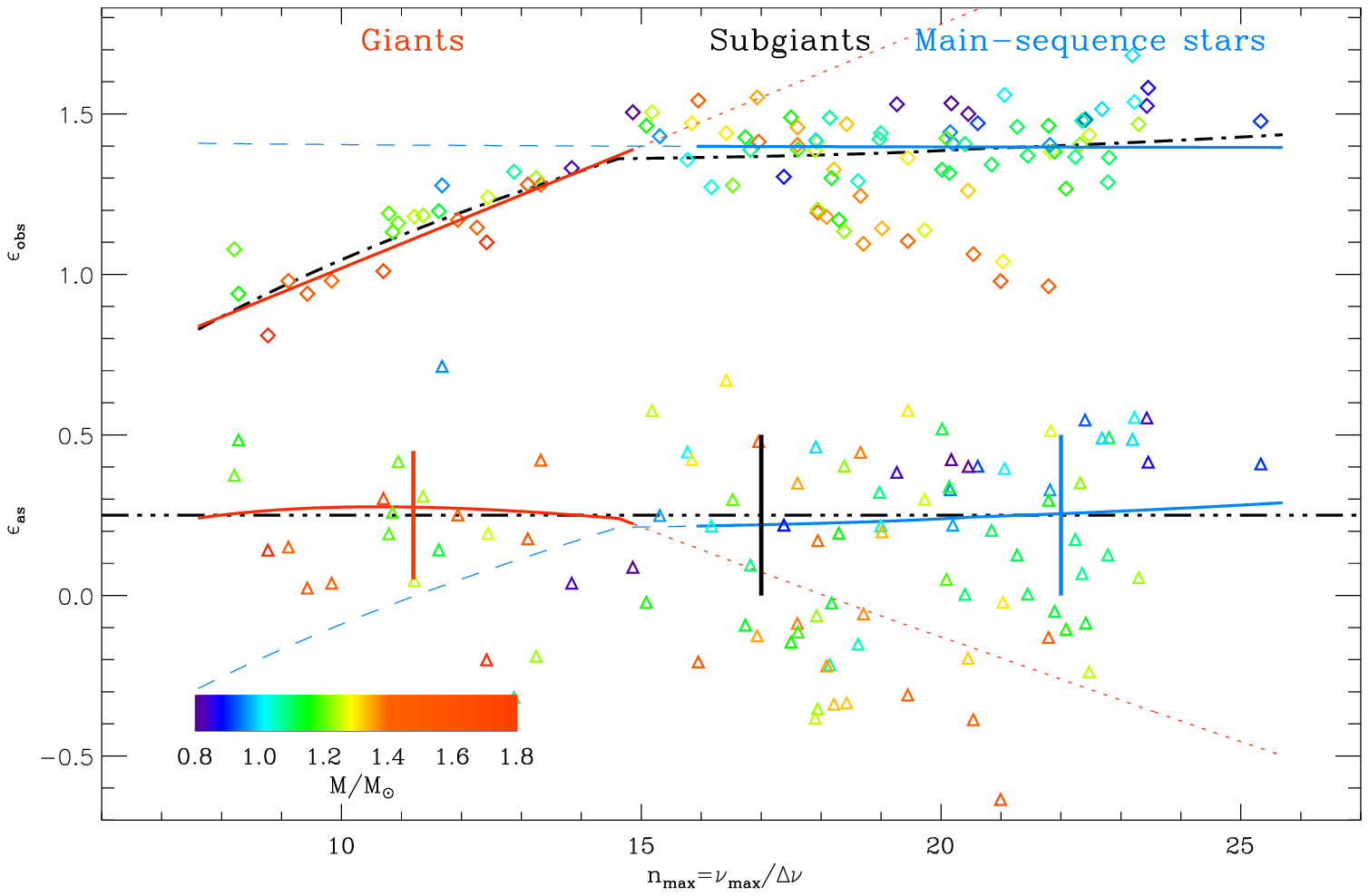}
\caption{Same as Fig.~\ref{asymp_second}, for the observed
(diamonds) and asymptotic (triangles)  offsets. Both parameters
are fitted, with dotted lines in the red giant regime and dashed
lines in the main-sequence regime; thicker lines indicate the
domain of validity of the fits. The triple-dot-dashed line
represents the Tassoul value $\epsas=1/4$, and the dot-dashed line
is the model of $\epsobs$ (varying with $\log\Dnuobs$ in the red
giant regime, and constant for less-evolved stars).
\label{asymp_eps}}
\end{figure*}

\begin{figure*}
\includegraphics[width=16cm]{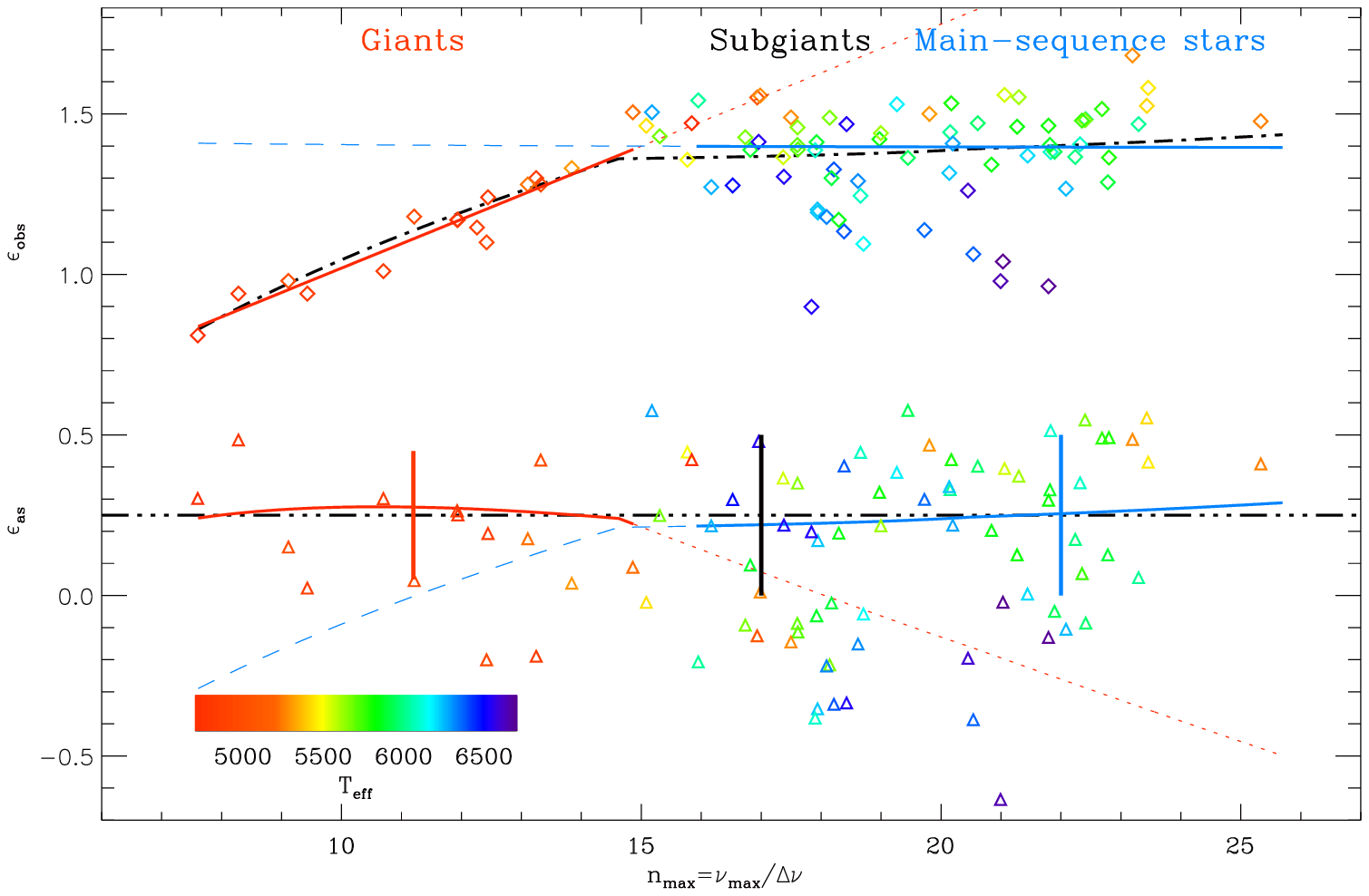}
\caption{Same as Fig.~\ref{asymp_eps}, with a color code depending
on the effective temperature. \label{asymp_eps_T}}
\end{figure*}

\subsection{Observed and asymptotic offsets\label{epsilon}}

The offsets $\epsobs$ and $\epsas$ are plotted in
Fig.~\ref{asymp_eps}. The fit of $\epsobs$, initially given by
\cite{2011A&A...525L...9M} for red giants and updated by
\cite{2012ApJ...757..190C}, is prolonged to main-sequence stars
with a nearly constant fit at $\epsobs{}\ind{,MS} \simeq 1.4$. We
based this fit on low-mass G stars in order to avoid the more
complex spectra of F stars that can be affected by the HD 49933
misidentification syndrome
\citep{2008A&A...488..705A,2009A&A...507L..13B}. The spread of
$\epsas$ is large since the propagation of the uncertainties from
$\alfa$ to $\epsas$ yields a large uncertainty:
$\delta\epsas\simeq 2 \delta\alfa / \alfa \simeq 0.4$.

We note in particular that the two different regimes seen for
$\alfa$ correspond to the different variations of $\epsobs$ with
stellar evolution. The comparison between $\epsas$ and $\epsobs$
allows us to derive significant features. The regime where
$\epsobs$ does not change with $\Dnuobs$ coincides with the regime
where the curvature evolves with $\nmax^{-2}$. The correction
provided by Eq.~\refeq{eqeps0} then mainly corresponds to the
constant term $2\,\coefaMS$. As a consequence, for low-mass
main-sequence and subgiant stars the asymptotic value is very
close to 1/4 \citep[Eq.~\refeq{surface}, as found
by][]{1980ApJS...43..469T}. In the red-giant regime, the curvature
varying as $\nmax^{-1}$ provides a variable correction, so that
$\epsas$ is close to 1/4 also even if $\epsobs$ varies with
$\Dnuobs$.

\subsection{Mass dependence}

For low-mass stars,  the fact that we find a mean value
$\langle\epsas\rangle$ of about 1/4 suggests that the asymptotic
expansion is valid for describing solar-like oscillation spectra
in a coherent way. This validity is also confirmed for red giants.
For those stars, we may assume that $\epsas \equiv 1/4$. Having
the observed value $\epsobs$ much larger than $\epsas$ can be seen
as an artefact of the curvature $\alfa$ introduced by the use of
$\Dnuobs$:
\begin{equation}
  \epsobs \simeq {1\over 4} \left( 1+ \nmax \displaystyle{\alfa\over 2}\right) + \nmax^2
  \alfa.
\end{equation}
However, we note in Fig.~\ref{asymp_eps} a clear gradient of
$\epsas$ with the stellar mass: high-mass stars have in general
lower $\epsas$ than low-mass stars, similar to what is observed
for $\epsobs$. In massive stars, the curvature is more pronounced;
$\epsobs$ and $\epsas$ are lower than in low-mass stars. As a
consequence, the asymptotic value $\epsas$ cannot coincide with
1/4, as is approximately the case for low-mass stars. We tried to
reconcile this different behavior by taking into account a mass
dependence in the fit of the curvature. In fact, fitting the
higher curvature of high-mass stars would translate into a larger
correction, in absolute value, from $\epsobs$ to $\epsas$, so that
it cannot account for the difference.

We are therefore left with the conclusion that, contrary to
low-mass stars, the asymptotic expansion is less satisfactory for
describing the radial-mode oscillation spectra of high-mass stars.
At this stage, we may imagine that in fact some features are
superimposed on the asymptotic spectrum, such as signatures of
glitches with longer period than the radial-order range where the
radial modes are observed. Such long-period glitches can be due to
the convective core in main-sequence stars with a mass larger than
1.2\,$M_\odot$; they are discussed in Section \ref{limitation}.

\section{Discussion\label{discussion}}

We explore here some consequences of fitting the observed
oscillation spectra with the exact asymptotic relation. First, we
examine the possible limitations of this relation as defined by
\cite{1980ApJS...43..469T}. Since the observed and asymptotic
values of the large separation differ, directly extracting the
stellar radius or mass from the measured value $\Dnuobs$ induces a
non-negligible bias. Thus, we revisit the scaling relations which
provide estimates of the stellar mass and radius. In a next step,
we investigate the meaning of the term $\epsobs$. We finally
discuss the consequence of having $\epsas$ exactly equal to 1/4.
This assumption would make the asymptotic expansion useful for
analysing acoustic glitches. Demonstrating that radial modes are
in all cases based on $\epsas \equiv 1/4$ will require some
modeling, which is beyond the scope of this paper.

\subsection{Contribution of the glitches\label{limitation}}

Solar and stellar oscillation spectra show that the asymptotic
expansion is not enough for describing the low-degree oscillation
pattern. Acoustic glitches yield significant modulation
\citep[e.g.,][and references therein]{2012A&A...540A..31M}.

Defining the global curvature is not an easy task, since the
radial oscillation is modulated by the glitches. With the
asymptotic description of the signature of the glitch proposed by
\cite{1993A&A...274..595P}, the asymptotic modulation adds a
contribution $\delta \nu_{n,0}$ to the eigenfrequency $\nu_{n,0}$
defined by Eqs. \refeq{asymp_ori} and \refeq{tassoul_as}, which
can be written at first order:
\begin{equation}\label{modulation}
    {\delta \nu_{n,0} \over \Dnuas} = \beta \sin 2\pi{n-\ng\over
    \Ng},
\end{equation}
where $\beta$ measures the amplitude of the glitch, $\ng$ its
phase, and $\Ng$ its period. This period varies as the ratio of
the stellar acoustic radius divided by the acoustic radius at the
discontinuity. Hence, deep glitches induce long-period modulation,
whereas glitches in the upper stellar envelope have short periods.
The phase $\ng$ has no simple expression. The \emph{observed}
large separations vary approximately as
\begin{equation}\label{modulationdnu}
    {\Delta\nu_{n,0} \over \Dnuobs} \simeq 1 + \alfa \left(n-\nmax\right) +
    {2\pi\beta\over \Ng} \cos 2\pi{n-\ng\over \Ng} ,
\end{equation}
according to the derivation of Eq.~\refeq{modulation}. In the
literature, we see typical values of $\Ng$ in the range 6 -- 12,
or even larger if the cause of the glitch is located in deep
layers. The amplitude of the modulation represents a few percent
of $\Dnuobs$, so that the modulation term ${2\pi\beta/ \Ng}$ can
greatly exceed the curvature $\alfa$. This explains the noisy
aspect of Figs.~\ref{asymp_second} to \ref{asymp_eps}, which is
due to larger spreads than the mean curvature. From $\epsas=1/4$,
one should get ${\epsobs}{}\ind{,as} \simeq 1.39$ for
main-sequence stars. If the observed value ${\epsobs}$ differs
from the expected asymptotic observed ${\epsobs}{}\ind{,as}$, then
one has to suppose that glitches explain the difference. The
departure from $\epsas=1/4$ of main-sequence stars with a larger
mass than 1.2\,$M_\odot$ is certainly related to the influence of
their convective core. If the period is long enough compared to
the number of observable modes, it can translate into an apparent
frequency offset, which is interpreted as an offset in $\epsobs$.
Similarly, glitches due to the high contrast density between the
core and the envelope in red giants, with a different phase
according to the evolutionary status, might modify differently the
curvature of their oscillation spectra. These hypotheses will be
tested in a forthcoming work.

\subsection{Scaling relations revisited\label{scaling}}

The importance of the measurements of $\Dnuobs$ and $\numax$ is
emphasized by their ability to provide relevant estimates of the
stellar mass and radius:
\begin{equation}
  {\Robs \over\Rs}  = \left({\numax \over \numaxs}\right) \
     \left({\Dnuobs \over \dnus}\right)^{-2}
     \left({\Teff \over \Ts}\right)^{1/2}, \label{scalingR}
\end{equation}
\begin{equation}
  {\Mobs\over\Ms} = \left({\numax \over \numaxs}\right)^{3}
     \left({\Dnuobs \over \dnus}\right)^{-4} \left({\Teff \over \Ts}\right)^{3/2} . \label{scalingM}
\end{equation}
The solar values chosen as references are not fixed uniformly in
the literature. Usually, internal calibration is ensured by the
analysis of the solar low-degree oscillation spectrum with the
same tool used for the asteroseismic spectra. Therefore, we used
$\dnus = 135.5\,\mu$Hz and $\numaxs = 3050\,\mu$Hz. One can find
significantly different values, e.g., $\dnus = 134.9\,\mu$Hz and
$\numaxs = 3120\,\mu$Hz \citep{2010A&A...522A...1K}. We will see
later that this diversity is not an issue if coherence and proper
calibration are ensured.

These relations rely on the definition of the large separation,
which scales as the square root of the mean stellar density
\citep{1917Obs....40..290E}, and $\numax$, which scales as the
cutoff acoustic frequency \citep{2011A&A...530A.142B}. From the
observed value $\Dnuobs$, one gets biased estimates $R(\Dnuobs)$
and  $M(\Dnuobs)$, since $\Dnuobs$ is underestimated when compared
to $\Dnuas$. Taking into account the correction provided by
Eq.~\refeq{eqdnu}, the corrected values are
\begin{equation}\label{corR}
  \Ras \simeq  \left({ 1 - 2\zeta} \right) \Robs \hbox{ and } \Mas \simeq  \left({ 1 - 4 \zeta} \right)
  \Mobs,
\end{equation}
with $\zeta$ defined by Eqs.~\refeq{corM_MS} or \refeq{corM_RGB},
depending on the regime. The amplitude of the correction $\zeta$
can be as high as 3.8\,\%, but one must take into account the fact
that scaling relations are calibrated on the Sun, so that one has
to deduce the solar correction $\zeta_\odot \simeq 2.6\,$\%. As a
result, for subgiants and main-sequence stars, the
\emph{systematic} negative correction reaches about 5\,\% for the
seismic estimate of the mass and about 2.5\,\% for the seismic
estimate of the radius. The absolute corrections are maximum in
the red giant regime. This justifies the correcting factors
introduced for deriving masses and radii for CoRoT red giants
\citep[Eqs. (9) and (10) of][]{2010A&A...517A..22M}, obtained by
comparison with the modeling of red giants chosen as reference.

We checked the bias in an independent way by comparing the stellar
masses obtained by modeling (sublist of \nmodeles\ stars in Table
\ref{valeurs}) with the seismic masses. These seismic estimates
appeared to be systematically 7.5\,\% larger and had to be
corrected. The comparison of the stellar radii yielded the same
conclusion: the mean bias was of the order 2.5\,\%. Furthermore,
we noted that the discrepancy increased significantly when the
stellar radius increased, as expected from Eqs.~\refeq{corR},
\refeq{corM_MS} and \refeq{corM_RGB}, since an increasing radius
corresponds to a decreasing value of $\nmax$.

This implies that  previous work based on the uncorrected scaling
relations might be improved. This concerns ensemble asteroseismic
results \citep[e.g.,][]{2011MNRAS.415.3539V,2011ApJ...743..143H}
or Galactic population analysis based on distance scaling
\citep{2009A&A...503L..21M}. This necessary systematic correction
must be prepared by an accurate calibration of the scaling
relations, taking into account the evolutionary status of the star
\citep{2011ApJ...743..161W,2012MNRAS.419.2077M}, cluster
properties \citep{2012MNRAS.419.2077M}, or independent
interferometric measurements \citep{2012ApJ...757...99S}.

At this stage, we suggest deriving the estimates of the stellar
mass and radius from a new set of equations based on $\Dnuas$ or
$\Dnuobs$:
\begin{equation}
  {\Rsis \over\Rs}  = \left({\numax \over \numaxas}\right) \
     \left({\Dnu \over \dnusas}\right)^{-2}
     \left({\Teff \over \Ts}\right)^{1/2}, \label{scalingRas}
\end{equation}
\begin{equation}
  {\Msis\over\Ms} = \left({\numax \over \numaxas}\right)^{3}
     \left({\Dnu \over \dnusas}\right)^{-4} \left({\Teff \over \Ts}\right)^{3/2} , \label{scalingMas}
\end{equation}
with new calibrated references $\numaxas = 3104\,\mu$Hz and
$\dnusas = 138.8\,\mu$Hz based on the comparison with the models
of Table~\ref{valeurs}. In case $\Dnuobs$ is used, corrections
provided by Eq.~\refeq{corR} must be applied; no correction is
needed with the use of $\Dnuas$. As expected from
\cite{2011ApJ...743..161W}, we noted that the quality of the
estimates decreased for effective temperatures higher than 6500\,K
or lower than 5000\,K. We therefore limited the calibration to
stars with a mass less than 1.3\,$M_\odot$. The accuracy of the
fit for these stars is about 8\,\% for the mass and 4\,\% for the
radius. Scaling relations are less accurate for  F stars with
higher mass and effective temperatures and for red giants. For
those stars, the performance of the scaling relations is degraded
by a factor 2.

\subsection{Surface effect?}

\subsubsection{Interpretation of $\epsobs$}

We note that the scaling of $\epsobs$ for red giants as a function
of $\log \Dnu$ implicitly or explicitly presented by many authors
\citep{2010ApJ...723.1607H,2011A&A...525L...9M,2012A&A...541A..51K,2012ApJ...757..190C}
can in fact be explained by the simple consequence of the
difference between $\Dnuobs$ and $\Dnuas$. Apart from the large
spread, we derived that the mean value of $\epsas$, indicated by
the dashed line in Fig.~\ref{asymp_eps}, is nearly a constant.

Following \cite{2012ApJ...751L..36W}, we examined how $\epsobs$
varies with effective temperature (Fig.~\ref{asymp_eps_T}).
Unsurprisingly, we see the same trend for subgiants and
main-sequence stars. Furthermore, there is a clear indication that
a similar gradient is present in red giants. However, there is no
similar trend for $\epsas$. This suggests that the physical reason
for explaining the gradient of $\epsobs$ with temperature is, in
fact, firstly related to the stellar mass and not to the effective
temperature.

The fact that $\epsobs$ is very different from $\epsas=1/4$ and
varies with the large separation suggests that $\epsobs$ cannot be
seen solely as an offset relating the surface properties
\citep[e.g.,][]{2012ApJ...751L..36W}. In other words, $\epsobs$
should not be interpreted as a surface parameter, since its
properties are closely related to the way the large separation is
determined. Its value is also severely affected by the glitches.
Any modulation showing a large period (in radial order) will
translate into an additional offset that will be mixed with
$\epsobs$.

In parallel, the mass dependence in $\epsas$ also indicates that
it depends on more than surface properties. Examining the exact
dependence of the offset $\epsas$ can be done by identifying it
with the phase shift given by the eigenfrequency equation
\citep{2000MNRAS.317..141R,2001MNRAS.322...85R}.

\subsubsection{Contribution of the upper atmosphere}

The uppermost part of the stellar atmosphere contributes to the
slight modification of the oscillation spectrum, since the level
of reflection of a pressure wave depends on its frequency
\citep[e.g.,][]{1994A&A...291.1019M}: the higher the frequency,
the higher the level of reflection; the larger the resonant
cavity, the smaller the apparent large separation. Hence, this
effect gives rise to apparent variation of the observed large
separations varying in the opposite direction compared to the
effect demonstrated in this work. This effect was not considered
in this work but must also be corrected for. It can be modeled
when the contribution of photospheric layers is taken into
account. For subgiants and main-sequence stars, its magnitude is
of about $-0.04\, \Dnu$ \citep{2012ApJ...749..152M}, hence much
smaller than the correction from $\Dnuobs$ to $\Dnuas$, which is
of about $+1.14\,\Dnu$ (Eq.~\refeq{fit_alpha_MS}).

\subsection{A generic asymptotic relation}

We now make the assumption that $\epsas$ is strictly equal to 1/4
for all solar-like oscillation patterns of low-mass stars and
explore the consequences.

\subsubsection{Subgiants and main-sequence stars}

The different scalings between observed and asymptotic seismic
parameters have indicated a dependence of the observed curvature
close to $\nmax^{-2}$ for subgiants and main-sequence stars. As a
result, it is possible to write the asymptotic relation for radial
modes (Eq.~\refeq{tassoul_as}) as
\begin{eqnarray}
\left.\nu_{\np,0}\right|\ind{MS} &=& \left(n  + \epsas +
\coefaMS\  {\nmax\over n} \right) \; \Dnuas \label{tassoul_MS} \\
  &\simeq& \left(n  + {1\over 4} +
\coefMS\  {\nmax\over n} \right) \; \Dnuas \\
  &\simeq& \left(n  + {1\over 4} +
{12.8\over n} \left({M \over\Ms}{\Rs \over R}{\Ts \over
\Teff}\right)^{1/2} \right) \; \Dnuas .
\end{eqnarray}
This underlines the large similarity of stellar interiors. With
such a development, the mean dimensionless value of the
second-order term is $\coefaMS$, since the ratio $\nmax / n$ is
close to 1 and does not vary with $\nmax$. However, compared to
the dominant term in $n$, its relative value decreases when
$\nmax$ increases. In absolute value, the second-order term scales
as $\Dnuas\, \numax / \nu$, so that the dimensionless term $d_1$
of Eq.~\ref{asymp_ori} is proportional to $\numax/\Dnuas$, with a
combined contribution of the stellar mean density ($\Dnuas$ term)
and acoustic cutoff frequency ($\numax$, hence $\nu\ind{c}$
contribution, \cite{2011A&A...530A.142B}).

\subsubsection{Red giants}

For red giants,  the asymptotic relation reduces to
\begin{eqnarray}
\left.\nu_{\np,0}\right|\ind{RG} &=& \left(n  + \epsas +
\coefa{}'{}\ind{RG} {\nmax^2\over n }  \right)\; \Dnuas \label{tassoul_RG} \\
  &\simeq& \left(n  + {1\over 4} +
0.037\  {\nmax^2\over n} \right) \; \Dnuas \\
  &\simeq& \left(n  + {1\over 4} +
{18.3\over n} \left({M \over\Ms}{\Rs \over R}{\Ts \over
\Teff}\right) \right) \; \Dnuas ,
\end{eqnarray}
with $ \coefa{}'{}\ind{RG} =  \coefaRG / (1+ \coefaRG)$. The
relationship given by Eq.~\refeq{fit_alpha_geantes} ensures that
the relative importance of the second-order term saturates when
$\nmax$ decreases. Its relative influence compared to the radial
order is $\coefa{}'{}\ind{RG}$, of about 4\,\%. Contrary to
less-evolved stars, the second-order term for red giants is in
fact proportional to $\numax^2 / \nu$ and is predominantly
governed by the surface gravity.

\subsubsection{Measuring the glitches\label{glitch}}

\begin{figure}
\includegraphics[width=8.8cm]{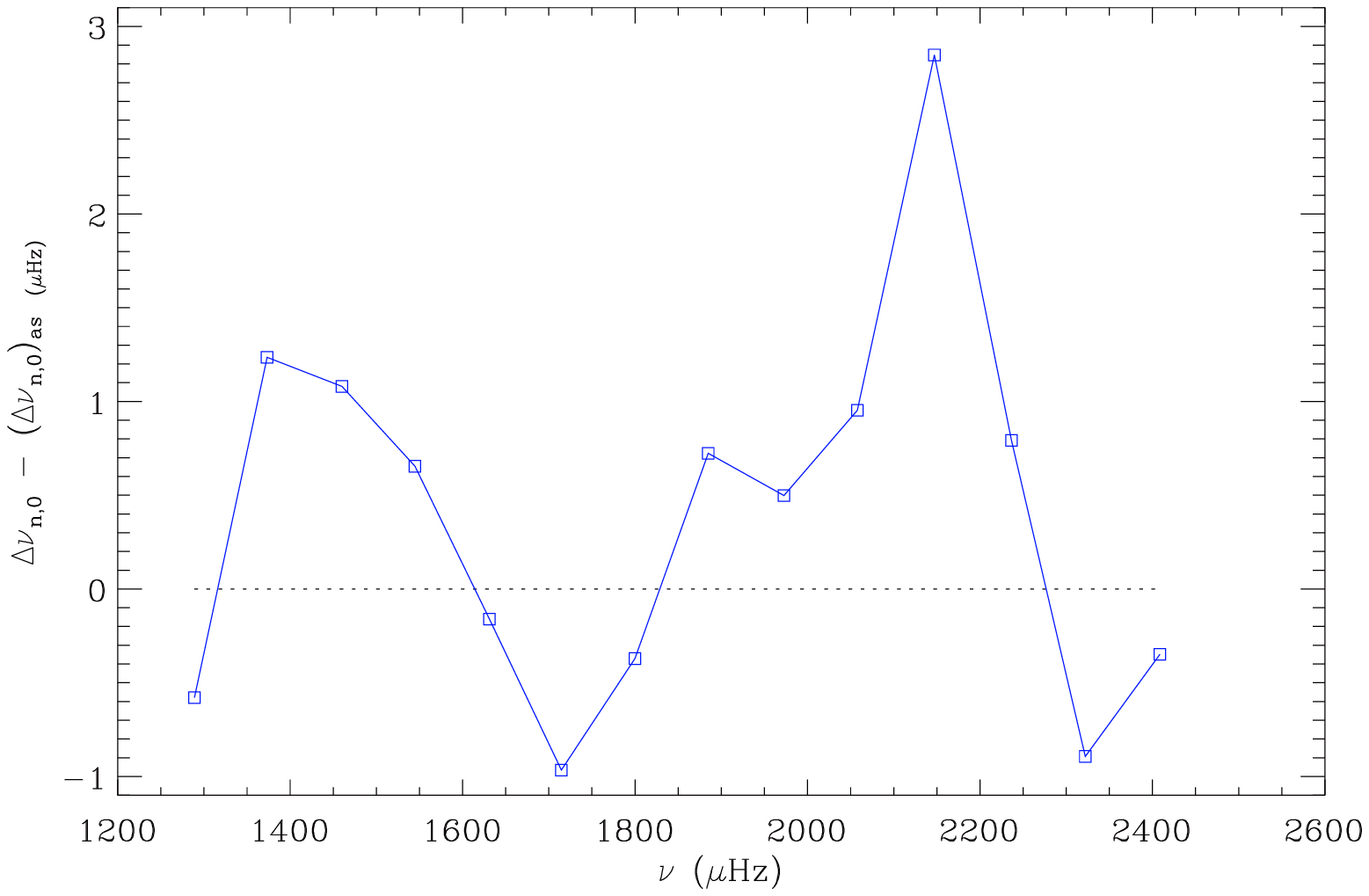}
\caption{Variation of the large separation difference
$\Delta\nu_{n+1,0} - (\Delta\nu_{n,0})\ind{as}$ as a function of
$\nu_{n,0}$ for the star HD\,49933.  \label{fig7}}
\end{figure}

These discrepancies to the generic relations express, in fact, the
variety of stars. In each regime, Eqs.~\refeq{tassoul_MS} and
\refeq{tassoul_RG} provide a reference for the oscillation
pattern, and Eq. \refeq{modulation} indicates the small departure
due to the specific interior properties of a star. Asteroseismic
inversion could not operate if all oscillation spectra were
degenerate and exactly similar to the mean asymptotic spectrum
depicted by Eq.~\refeq{tassoul_MS} and \refeq{tassoul_RG}).
Conversely, if one assumes that the asymptotic relation provides a
reliable reference case for a given stellar model represented by
its large separation, then comparing an observed spectrum to the
mean spectrum expected at $\Dnu$ may provide a way to determine
the glitches. In other words, glitches may correspond to observed
modulation after subtraction of the mean curvature. We show this
in an example (Fig.~\ref{fig7}), where we subtracted the mean
asymptotic slope derived from Eq.~\refeq{fit_alpha_MS} from the
local large separations $\nu_{n+1,0}-\nu_{n,0}$.

\section{Conclusion}

We have addressed some consequences of the expected difference
between the observed and asymptotic values of the large
separation. We derived the curvature of the radial-mode
oscillation pattern in a large set of solar-like oscillation
spectra from the variation in frequency of the spacings between
consecutive radial orders. We then proposed a simple model to
represent the observed spectra with the exact asymptotic form.
Despite the spread of the data, the ability of the Tassoul
asymptotic expansion to account for the solar-like oscillation
spectra is confirmed, since we have demonstrated coherence between
the observed and asymptotic parameters. Two regimes have been
identified: one corresponds to the subgiants and main-sequence
stars, the other to red giants. These regimes explain the
variation of the observed offset $\epsobs$ with the large
separation.

\paragraph*{Curvature and second-order effect:}

\ We have verified that the curvature observed in the \'echelle
diagram of solar-like oscillation spectra corresponds to the
second-order term of the Tassoul equation. We have shown that the
curvature scales approximately as $(\Dnu / \numax )^{2}$ for
subgiants and main-sequence stars. Its behavior changes in the red
giant regime, where it varies approximately as $(\Dnu / \numax)$
for red giants.

\paragraph*{Large separation and scaling relations:}

\ As the ratio $\Dnuas / \Dnuobs$ changes along the stellar
evolution (from about 2\,\% for a low-mass dwarf to 4\,\% for a
giant), scaling relations must be corrected to avoid a systematic
overestimate of the seismic proxies. Corrections are proposed for
the stellar radius and mass, which avoid a bias of 2.5\,\% for $R$
and 5\,\% for $M$. The corrected and calibrated scaling relations
then provide an estimate of $R$ and $M$ with 1-$\sigma$
uncertainties of, respectively, 4 and 8\,\% for low-mass stars.

\paragraph*{Offsets:}

\ The observed values $\epsobs$ are affected by the definition and
the measurement of the large separation. Their spread is amplified
by all glitches affecting solar-like oscillation spectra. We made
clear that the variation of $\epsobs$ with stellar evolution is
mainly an artefact due to the use of $\Dnuobs$ instead of $\Dnuas$
in the asymptotic relation. This work shows that the asymptotic
value $\epsas$ is a small constant which does vary much throughout
stellar evolution for low-mass stars and red giants. It is very
close to the value 1/4 derived from the asymptotic expansion for
low-mass stars.

\paragraph*{Generic asymptotic relation:}

According to this, we have established a generic form of the
asymptotic relation of radial modes in low-mass stars. The mean
oscillation pattern based on this relation can serve as a
reference for rapidly analyzing a large amount of data, for
identifying unambiguously an oscillation pattern, and for
determining acoustic glitches. Departure from the generic
asymptotic relation is expected for main-sequence stars with a
convective core and for red giants, according to the asymptotic
expansion of structure discontinuities.
\\

Last but not least, this work implies that all the quantitative
conclusions of previous analyses that have created confusion
between $\Dnuas$ and $\Dnuobs$ have to be reconsidered. The
observed values derived from the oscillation spectra have to be
translated into asymptotic values before any physical analysis. To
avoid confusion, it is also necessary to specify which value of
the large separation is considered. A coherent notation for the
observed value of the large separation at $\numax$ should be
$\Dnu\ind{max}$.

%

\bibliographystyle{aa} 
\bibliography{biblio_mesure}

\longtab{1}{\tiny
\begin{longtable}{lcrrrrrrrrrrl}
\caption{Stellar parameters\label{valeurs}}\\
\hline\hline
Star$^{(a)}$&$\nmax$ &$\Dnuobs$ &$\Dnuas^{(b)}$ &$\numax$& $\epsobs$ &$\alfa$ &$\Teff^{(c)}$& $\Mmod^{(d)}$ & $\Msis^{(e)}$ & $\Rmod^{(d)}$ & $\Rsis^{(e)}$ & Ref.$^{(f)}$\\
 & & ($\mu$Hz)&($\mu$Hz)&($\mu$Hz)& & $\times 10^3$ & (K) & $(M_\odot)$& $(M_\odot)$ & $(R_\odot)$& $ (R_\odot)$ & \\
\hline
\endfirsthead
\caption{continued.}\\
\hline\hline
Star$^{(a)}$&$\nmax$ &$\Dnuobs$ &$\Dnuas^{(b)}$ &$\numax$& $\epsobs$ &$\alfa$ &$\Teff^{(c)}$& $\Mmod^{(d)}$ & $\Msis^{(e)}$ & $\Rmod^{(d)}$ & $\Rsis^{(e)}$ & Ref.$^{(f)}$\\
 & & ($\mu$Hz)&($\mu$Hz)&($\mu$Hz)& & $\times 10^3$ & (K) & $(M_\odot)$& $(M_\odot)$ & $(R_\odot)$& $ (R_\odot)$ & \\
\hline
\endhead
\hline
\endfoot
 HD 181907    &  8.2 & 3.41 & 3.55 & 28.0 & 1.08 &  10.2 &4790 & 1.20 & 1.29 &12.20 & 12.6 &           2010Car, 2010Mig  \\
 KIC 4044238  &  8.3 & 4.07 & 4.23 & 33.7 & 0.94 &   6.5 &4800 &      & 1.11 &      & 10.6 &                     2012Mos \\
 HD 50890     &  8.8 & 1.71 & 1.78 & 15.0 & 0.81 &   8.6 &4670 & 4.20 & 3.03 &29.90 & 26.4 &                     2012Bau \\
 KIC 5000307  &  9.1 & 4.74 & 4.93 & 43.2 & 0.98 &   9.9 &4992 &      & 1.35 &      & 10.2 &                     2012Mos \\
 KIC 9332840  &  9.4 & 4.39 & 4.57 & 41.4 & 0.94 &  10.3 &4847 &      & 1.55 &      & 11.3 &                     2012Mos \\
 KIC 4770846  &  9.8 & 5.48 & 5.70 & 53.9 & 0.98 &   9.7 &4801 &      & 1.39 &      & 9.37 &                     2012Mo2 \\
 KIC 2013502  & 10.7 & 5.72 & 5.95 & 61.2 & 1.01 &   6.1 &4835 &      & 1.73 &      & 9.80 &                     2012Mos \\
 KIC 10866415 & 10.8 & 8.75 & 9.10 & 94.4 & 1.19 &   8.5 &4812 &      & 1.15 &      & 6.44 &                     2012Mo2 \\
 KIC 11550492 & 10.9 & 8.70 & 9.05 & 94.4 & 1.13 &   7.3 &4723 &      & 1.14 &      & 6.46 &                     2012Mo2 \\
 KIC 9574650  & 10.9 & 9.64 &10.03 &  105 & 1.16 &   6.1 &5015 &      & 1.16 &      & 6.06 &                     2012Mo2 \\
 KIC 3744043  & 11.2 & 9.90 &10.30 &  111 & 1.18 &   9.0 &4994 &      & 1.21 &      & 6.03 &                     2012Mos \\
 KIC 9267654  & 11.4 & 10.4 & 10.8 &  117 & 1.18 &   6.7 &4965 &      & 1.19 &      & 5.83 &                     2012Mo2 \\
 KIC 6144777  & 11.6 & 11.0 & 11.5 &  128 & 1.20 &   7.8 &4657 &      & 1.09 &      & 5.42 &                     2012Mo2 \\
 KIC 5858947  & 11.7 & 14.5 & 15.1 &  169 & 1.28 &   4.0 &4977 &      & 0.92 &      & 4.27 &                     2012Mo2 \\
 KIC 6928997  & 11.9 & 10.1 & 10.5 &  120 & 1.17 &   6.4 &4800 &      & 1.35 &      & 6.20 &            2011Bec, 2012Mos \\
 KIC 11618103 & 12.3 & 9.38 & 9.76 &  115 & 1.15 &  10.0 &4870 & 1.45 & 1.60 & 6.60 & 6.87 &                     2011Jia \\
 KIC 8378462  & 12.4 & 7.27 & 7.56 & 90.3 & 1.10 &   8.5 &4962 &      & 2.21 &      & 9.06 &                     2012Mos \\
 KIC 12008916 & 12.4 & 12.9 & 13.4 &  159 & 1.24 &   6.7 &4830 & 1.26 & 1.21 & 5.18 & 5.07 &                     2012Bec \\
 KIC 9882316  & 13.1 & 13.7 & 14.2 &  179 & 1.28 &   6.4 &5228 &      & 1.49 &      & 5.22 &                     2012Mos \\
 KIC 5356201  & 13.2 & 15.8 & 16.5 &  209 & 1.30 &   8.6 &4840 & 1.23 & 1.19 & 4.47 & 4.38 &                     2012Bec \\
 KIC 8366239  & 13.3 & 13.7 & 14.2 &  182 & 1.28 &   4.8 &4980 & 1.49 & 1.46 & 5.30 & 5.18 &                     2012Bec \\
 KIC 7341231  & 13.8 & 28.8 & 30.0 &  399 & 1.33 &   6.7 &5300 & 0.83 & 0.85 & 2.62 & 2.63 &                   2012Deh   \\
 KIC 11717120 & 14.9 & 37.3 & 38.8 &  555 & 1.50 &   6.4 &5150 &      & 0.78 &      & 2.15 &            2012App, 2012Bru \\
 KIC  9574283 & 15.1 & 29.7 & 30.9 &  448 & 1.46 &   6.5 &5440 &      & 1.11 &      & 2.82 &                     2012App \\
 KIC  8026226 & 15.2 & 34.3 & 35.6 &  520 & 1.50 &   4.0 &6230 &      & 1.21 &      & 2.64 &            2012App, 2012Bru \\
 KIC  5607242 & 15.3 & 39.8 & 41.4 &  610 & 1.43 &   5.0 &5680 &      & 0.93 &      & 2.18 &                     2012App \\
 KIC  8702606 & 15.8 & 39.7 & 41.2 &  626 & 1.36 &   3.6 &5540 &      & 0.99 &      & 2.23 &            2012App, 2012Bru \\
 KIC  4351319 & 15.8 & 24.4 & 25.4 &  387 & 1.47 &   4.1 &4700 & 1.30 & 1.27 & 3.40 & 3.36 &                     2011diM \\
 KIC 11771760 & 16.0 & 31.7 & 32.9 &  505 & 1.54 &   6.9 &6030 &      & 1.46 &      & 2.96 &                 2012App    \\
 KIC  7976303 & 16.2 & 51.1 & 53.0 &  826 & 1.27 &   4.0 &6260 &      & 1.00 &      & 1.90 &                     2012App \\
 HD 182736    & 16.4 & 34.6 & 35.9 &  568 & 1.44 &   2.8 &5261 & 1.30 & 1.19 & 2.70 & 2.61 &                     2012Hub \\
 KIC 10909629 & 16.5 & 49.2 & 51.0 &  813 & 1.28 &   3.5 &6490 &      & 1.17 &      & 2.05 &                     2012App \\
 KIC 11414712 & 16.7 & 43.6 & 45.2 &  730 & 1.43 &   5.4 &5635 &      & 1.11 &      & 2.19 &            2012App, 2012Bru \\
 KIC  5955122 & 16.8 & 49.1 & 50.9 &  826 & 1.39 &   4.6 &5837 &      & 1.06 &      & 1.99 &            2012App, 2012Bru \\
 KIC  7799349 & 16.9 & 33.1 & 34.2 &  560 & 1.55 &   5.9 &5115 &      & 1.31 &      & 2.78 &            2012App, 2012Bru \\
 Procyon      & 17.0 & 54.1 & 56.0 &  918 & 1.41 &   3.2 &6550 & 1.46 & 1.17 & 2.04 & 1.93 &            2010Bed, 2008Mos \\
 KIC  1435467 & 17.4 & 79.6 & 82.4 & 1384 & 1.30 &   3.6 &6570 &      & 0.86 &      & 1.35 &            2012App, 2012Bru \\
 KIC 12508433 & 17.5 & 44.8 & 46.3 &  784 & 1.49 &   5.4 &5280 &      & 1.13 &      & 2.16 &                     2012App \\
 KIC 11193681 & 17.6 & 42.7 & 44.2 &  752 & 1.40 &   4.8 &5690 &      & 1.35 &      & 2.37 &                     2012App \\
 KIC 11395018 & 17.6 & 47.4 & 49.0 &  834 & 1.46 &   3.5 &5650 & 1.35 & 1.20 & 2.21 & 2.13 &            2011Mat, 2012Cre \\
 KIC 11026764 & 17.6 & 50.3 & 52.0 &  885 & 1.39 &   4.9 &5682 &      & 1.14 &      & 2.01 &   2011Cam, 2012App, 2012Bru \\
 HD 49385     & 17.9 & 55.5 & 57.4 &  994 & 1.39 &   5.6 &6095 & 1.25 & 1.21 & 1.94 & 1.92 &            2010Deh, 2011Deh \\
 KIC 11713510 & 17.9 & 68.9 & 71.3 & 1235 & 1.42 &   2.9 &5930 & 1.00 & 0.94 & 1.57 & 1.53 &                     2012Mat \\
 KIC 10920273 & 17.9 & 57.2 & 59.2 & 1026 & 1.41 &   4.6 &5880 & 1.23 & 1.12 & 1.88 & 1.83 &            2011Cam, 2012Cre \\
 KIC 10018963 & 17.9 & 55.1 & 56.9 &  988 & 1.20 &   4.9 &6020 &      & 1.21 &      & 1.93 &            2012App, 2012Bru \\
 KIC  3632418 & 17.9 & 60.4 & 62.4 & 1084 & 1.19 &   3.2 &6190 & 1.40 & 1.15 & 1.91 & 1.78 &   2012App, 2012Sil, 2012Bru \\
 KIC 10162436 & 18.1 & 55.5 & 57.3 & 1004 & 1.18 &   4.3 &6200 & 1.36 & 1.28 & 2.01 & 1.96 &   2012App, 2012Sil, 2012Bru \\
 KIC  8524425 & 18.1 & 59.4 & 61.4 & 1078 & 1.49 &   5.2 &5634 &      & 1.05 &      & 1.75 &            2012App, 2012Bru \\
 KIC  7747078 & 18.2 & 53.8 & 55.5 &  977 & 1.30 &   4.0 &5840 & 1.13 & 1.23 & 1.95 & 1.97 &   2012App, 2012Sil, 2012Bru \\
 KIC  7103006 & 18.2 & 58.9 & 60.8 & 1072 & 1.33 &   5.1 &6394 &      & 1.30 &      & 1.89 &            2012App, 2012Bru \\
 HD 169392    & 18.3 & 56.3 & 58.1 & 1030 & 1.17 &   2.9 &5850 & 1.15 & 1.20 & 1.88 & 1.90 &                     2012Ma2 \\
 KIC  9812850 & 18.4 & 64.5 & 66.6 & 1186 & 1.13 &   2.1 &6325 &      & 1.20 &      & 1.73 &            2012App, 2012Bru \\
 KIC 12317678 & 18.4 & 63.1 & 65.1 & 1162 & 1.47 &   5.4 &6540 &      & 1.30 &      & 1.80 &                     2012App \\
 KIC  8694723 & 18.6 & 74.3 & 76.7 & 1384 & 1.29 &   4.2 &6120 &      & 1.03 &      & 1.50 &            2012App, 2012Bru \\
 KIC  8228742 & 18.7 & 61.8 & 63.8 & 1153 & 1.25 &   2.3 &6042 & 1.38 & 1.22 & 1.85 & 1.79 &   2012App, 2012Sil, 2012Bru \\
 KIC 10273246 & 18.7 & 48.8 & 50.4 &  913 & 1.10 &   3.3 &6150 & 1.37 & 1.60 & 2.19 & 2.30 &            2011Cam, 2012Cre \\
 KIC  6933899 & 19.0 & 71.8 & 74.1 & 1362 & 1.42 &   3.0 &5860 &      & 1.06 &      & 1.55 &            2012App, 2012Bru \\
 KIC 11244118 & 19.0 & 71.2 & 73.4 & 1352 & 1.44 &   3.4 &5745 &      & 1.04 &      & 1.55 &            2012App, 2012Bru \\
 HD 179070    & 19.0 & 60.6 & 62.6 & 1153 & 1.14 &   2.6 &6131 & 1.34 & 1.35 & 1.86 & 1.88 &                     2012How \\
 KIC  9410862 & 19.3 &105.6 &108.9 & 2034 & 1.53 &   3.1 &6180 &      & 0.82 &      & 1.10 &                     2012App \\
 KIC 10355856 & 19.4 & 67.0 & 69.1 & 1303 & 1.10 &   3.8 &6350 &      & 1.38 &      & 1.77 &            2012App, 2012Bru \\
 KIC 12258514 & 19.4 & 74.5 & 76.8 & 1449 & 1.36 &   2.0 &5930 & 1.30 & 1.12 & 1.63 & 1.54 &   2012App, 2012Sil, 2012Bru \\
 KIC  7206837 & 19.7 & 78.9 & 81.3 & 1556 & 1.14 &   2.1 &6384 &      & 1.24 &      & 1.53 &            2012App, 2012Bru \\
 KIC 10516096 & 20.0 & 84.4 & 86.9 & 1689 & 1.33 &   2.0 &5940 & 1.12 & 1.09 & 1.42 & 1.40 &            2012Mat, 2012Bru \\
 KIC 7680114  & 20.1 & 84.9 & 87.4 & 1705 & 1.42 &   3.4 &5855 & 1.19 & 1.07 & 1.45 & 1.39 &            2012Mat, 2012Bru \\
 KIC 12009504 & 20.1 & 87.8 & 90.4 & 1768 & 1.32 &   2.4 &6065 &      & 1.10 &      & 1.37 &            2012App, 2012Bru \\
 KIC  6116048 & 20.2 &100.2 &103.2 & 2020 & 1.44 &   2.7 &5935 &      & 0.93 &      & 1.19 &            2012App, 2012Bru \\
 KIC  8760414 & 20.2 &116.4 &119.9 & 2349 & 1.53 &   2.7 &5787 & 0.81 & 0.78 & 1.02 & 1.01 &   2012Mat, 2012App, 2012Bru \\
 KIC 10963065 & 20.2 &102.6 &105.6 & 2071 & 1.41 &   2.9 &6060 &      & 0.95 &      & 1.18 &            2012App, 2012Bru \\
 KIC 3656476  & 20.4 & 93.3 & 96.1 & 1904 & 1.41 &   3.4 &5710 & 1.09 & 0.98 & 1.32 & 1.27 &            2012Mat, 2012Bru \\
 KIC 11081729 & 20.4 & 89.0 & 91.6 & 1820 & 1.26 &   3.5 &6630 &      & 1.30 &      & 1.44 &            2012App, 2012Bru \\
 HD 46375     & 20.5 &154.0 &158.5 & 3150 & 1.50 &   2.6 &5300 &      & 0.54 &      & 0.74 &                     2010Gau \\
 KIC  9139163 & 20.5 & 80.3 & 82.6 & 1649 & 1.06 &   3.5 &6400 & 1.40 & 1.38 & 1.57 & 1.57 & 2012App, 2012Sil, 2012Bru   \\
 KIC  9098294 & 20.6 &108.7 &111.9 & 2241 & 1.47 &   2.5 &5840 &      & 0.90 &      & 1.11 &            2012App, 2012Bru \\
 KIC 4914923  & 20.8 & 88.7 & 91.2 & 1848 & 1.34 &   2.6 &5905 & 1.10 & 1.16 & 1.37 & 1.39 &            2012Mat, 2012Bru \\
 HD 181420    & 21.0 & 74.9 & 77.1 & 1573 & 0.98 &   3.7 &6580 & 1.58 & 1.65 & 1.69 & 1.75 &            2009Bar, 2012Oze \\
 HD 49933     & 21.0 & 85.8 & 88.3 & 1805 & 1.04 &   2.4 &6780 & 1.30 & 1.52 & 1.41 & 1.55 &            2009Ben, 2012Ree \\
 KIC 6603624  & 21.1 &109.8 &112.9 & 2312 & 1.56 &   2.6 &5625 & 1.01 & 0.90 & 1.15 & 1.11 &   2012Mat, 2012App, 2012Bru \\
 16 Cyg A     & 21.3 &102.5 &105.4 & 2180 & 1.46 &   2.9 &5825 & 1.11 & 1.05 & 1.24 & 1.22 &                     2012Met \\
 KIC  8394589 & 21.4 &108.9 &112.0 & 2336 & 1.37 &   3.0 &6114 &      & 1.09 &      & 1.19 &            2012App, 2012Bru \\
 KIC 11253226 & 21.8 & 77.0 & 79.1 & 1678 & 0.96 &   2.3 &6605 & 1.46 & 1.82 & 1.63 & 1.77 & 2012App, 2012Sil, 2012Bru   \\
 \muArae      & 21.8 & 89.5 & 91.9 & 1950 & 1.46 &   2.4 &5813 & 1.14 & 1.29 & 1.36 & 1.43 &            2005Bou, 2005Baz \\
 HD 52265     & 21.8 & 98.5 &101.2 & 2150 & 1.38 &   1.8 &6100 & 1.27 & 1.27 & 1.34 & 1.34 &            2011Bal, 2012Giz \\
 KIC 6106415  & 21.9 &103.9 &106.7 & 2274 & 1.38 &   3.0 &5990 & 1.12 & 1.18 & 1.24 & 1.26 &            2012Mat, 2012Bru \\
 KIC 10454113 & 22.1 &104.7 &107.6 & 2313 & 1.27 &   2.8 &6120 & 1.16 & 1.24 & 1.25 & 1.27 &   2012App, 2012Sil, 2012Bru \\
 KIC  8379927 & 22.2 &120.0 &123.3 & 2669 & 1.37 &   2.4 &5990 &      & 1.07 &      & 1.11 &                     2012App \\
 KIC  9139151 & 22.3 &116.9 &120.1 & 2610 & 1.40 &   2.1 &6125 & 1.22 & 1.15 & 1.18 & 1.15 &   2012App, 2012Sil, 2012Bru \\
 16 Cyg B     & 22.4 &115.3 &118.4 & 2578 & 1.48 &   2.8 &5750 & 1.07 & 1.07 & 1.13 & 1.14 &                     2012Met \\
 KIC  9025370 & 22.4 &132.3 &135.9 & 2964 & 1.48 &   1.8 &5660 &      & 0.91 &      & 0.98 &                  2012App   \\
 KIC  3427720 & 22.4 &119.3 &122.5 & 2674 & 1.48 &   3.1 &6040 &      & 1.12 &      & 1.13 &            2012App, 2012Bru \\
 KIC 5184732  & 22.5 & 95.3 & 97.9 & 2142 & 1.43 &   3.3 &5840 & 1.25 & 1.34 & 1.36 & 1.39 &            2012Mat, 2012Bru \\
 Sun          & 22.7 &134.4 &138.0 & 3050 & 1.51 &   2.0 &5777 & 1.00 & 0.97 & 1.00 & 0.99 &                         Sun \\
 KIC 8379927  & 22.8 &120.4 &123.6 & 2743 & 1.29 &   2.2 &5900 & 1.09 & 1.13 & 1.11 & 1.12 &            2012Mat, 2012App \\
 \aCenA       & 22.8 &105.7 &108.5 & 2410 & 1.36 &   1.7 &5790 & 1.10 & 1.25 & 1.23 & 1.27 &           2004Bed, 2002The  \\
 KIC 8006161  & 23.2 &148.5 &152.3 & 3444 & 1.68 &   2.2 &5390 & 1.00 & 0.84 & 0.93 & 0.89 &   2012Mat, 2012App, 2012Bru \\
 18 Sco       & 23.2 &133.5 &136.9 & 3100 & 1.54 &   1.8 &5813 & 1.02 & 1.06 & 1.01 & 1.03 &            2011Baz, 2012Baz \\
 KIC 10644253 & 23.3 &123.0 &126.2 & 2866 & 1.47 &   2.6 &6030 &      & 1.22 &      & 1.14 &            2012App, 2012Bru \\
 KIC 11772920 & 23.4 &155.3 &159.3 & 3639 & 1.52 &   1.8 &5420 &      & 0.84 &      & 0.86 &                     2012App \\
 KIC  9955598 & 23.5 &152.6 &156.5 & 3579 & 1.58 &   2.1 &5410 &      & 0.85 &      & 0.88 &         2012App, 2012Bru   \\
 \aCenB       & 25.3 &161.4 &165.3 & 4090 & 1.48 &   1.6 &5260 & 0.91 & 0.98 & 0.86 & 0.88 &           2005Kje, 2002The  \\
\end{longtable}
\noindent \emph{(a)} Stars are sorted by increasing $\nmax$ values. \\
\emph{(b)} $\Dnuas$ is derived from this work.\\
\emph{(c)} $\Teff$ provided by \cite{2012MNRAS.423..122B}, when available.\\
\emph{(d)} $\Mmod$ and $\Rmod$ were obtained from modeling or from interferometry, not from seismic scaling relations. \\
\emph{(e)} $\Msis$ and $\Rsis$ are derived from the scaling relations Eqs.~\refeq{scalingRas} and \refeq{scalingMas}
with the new calibrations defined by this work. \\
\emph{(f)} References are defined by:\\
2002The = \cite{2002A&A...392L...9T}   \\
2004Bed = \cite{2004ApJ...614..380B}   \\
2005Baz = \cite{2005A&A...440..615B}   \\
2005Bou = \cite{2005A&A...440..609B}   \\
2005Kje = \cite{2005ApJ...635.1281K}   \\
2006Cat = \cite{2006AJ....132.2318C}   \\
2008Mos = \cite{2008A&A...478..197M}   \\
2009Bar = \cite{2009A&A...506...51B}   \\
2009Ben = \cite{2009A&A...507L..13B}   \\
2010Bed = \cite{2010ApJ...713..935B}   \\
2010Car = \cite{2010A&A...509A..73C}   \\
2010Deh = \cite{2010A&A...515A..87D}   \\
2010Gau = \cite{2010A&A...524A..47G}   \\
2010Mig = \cite{2010A&A...520L...6M}   \\
2011Bal = \cite{2011A&A...530A..97B}   \\
2011Baz = \cite{2011A&A...526L...4B}   \\
2011Bec = \cite{2011Sci...332..205B}   \\
2011Cam = \cite{2011A&A...534A...6C}   \\
2011Deh = \cite{2011A&A...535A..91D}   \\
2011diM = \cite{2011MNRAS.415.3783D}   \\
2011Jia = \cite{2011ApJ...742..120J}   \\
2011Mat = \cite{2011ApJ...733...95M}   \\
2012App = \cite{2012A&A...543A..54A}   \\
2012Bau = \cite{2012A&A...538A..73B}   \\
2012Baz = \cite{2012A&A...544A.106B}   \\
2012Bec = \cite{2012Natur.481...55B}   \\
2012Bru = \cite{2012MNRAS.423..122B} \\
2012Cre = \cite{2012A&A...537A.111C}   \\
2012Deh = \cite{2012ApJ...756...19D}   \\
2012Giz = Gizon et al. (2012)          \\
2012How = \cite{2012ApJ...746..123H}   \\
2012Hub = \cite{2012ApJ...760...32H}   \\
2012Mat = \cite{2012ApJ...749..152M}   \\
2012Ma2 = \cite{2013A&A...549A..12M}   \\
2012Met = \cite{2012ApJ...748L..10M}   \\
2012Mos = \cite{2012A&A...540A.143M}   \\
2012Mo2 = \cite{2012A&A...548A..10M}   \\
2012Oze = Ozel et al. (2012)           \\
2012Ree = \cite{2012A&A...539A..63R}   \\
2012Sil = \cite{2012ApJ...757...99S}   \\
}

\end{document}